\documentclass[twocolumn]{aastex631}

\def \HI{{\sc Hi}}
\usepackage{multirow}
\usepackage{graphicx}	% Including figure files
\usepackage{amsmath}	% Advanced maths commands
\usepackage{mathtools}
\usepackage{hyperref}
\usepackage[mathscr]{euscript}
\usepackage{nccmath}
\DeclareUnicodeCharacter{2212}{-}
\usepackage{tikz}
\usepackage{overpic}

\begin{document}

\title{{Investigating radio source population and spatial characteristics of diffuse Galactic synchrotron emission in the GAMA-23 field with uGMRT Band-3}}

\correspondingauthor{Rashmi Sagar}
\email{rashmisagar777@gmail.com, phd2101121003@iiti.ac.in}

\author[0009-0007-3170-4835]{Rashmi Sagar}
\affiliation{Department of Astronomy, Astrophysics and Space Engineering, Indian Institute of Technology Indore, Indore 452020, India}
\email[hide]{rashmisagar777@gmail.com}  
\email[hide]{phd2101121003@iiti.ac.in}  
\author[0000-0002-5333-1095]{Abhirup Datta}
\affiliation{Department of Astronomy, Astrophysics and Space Engineering, Indian Institute of Technology Indore, Indore 452020, India}
\email[hide]{abhirup.datta@iiti.ac.in}  
%\collaboration{20}{(AAS Journals Data Editors)}

\author{Aishrila Mazumder}
\affiliation{Jodrell Bank Centre for Astrophysics, Department of Physics and Astronomy, The University of Manchester, Manchester M13 9PL, UK}
\email[hide]{fakeemail1@google.com}

% \email[hide]{fakeemail1@google.com}  


\begin{abstract}
We present upgraded Giant Meterwave Radio Telescope (uGMRT) Band-3 observations of the Galaxy and Mass Assembly (GAMA)-23 field. The observations consist of a total of 33 hr spread over 50 pointings, corresponding to $\sim$\,56 minutes per pointing at the central frequency of 325\,MHz. The final image mosaicked from these pointings has a central off-source RMS noise of $109\,\mu\mathrm{Jy}\,\mathrm{beam}^{-1}$ and covers an area $65.31\,\text{deg}^2$ with a resolution of $15.8''$. This wide-area deep-field observation provides a radio source catalog of 5741 sources with flux densities ${\geq}5{\sigma}$. We present complementary uGMRT observations overlapping with the high-frequency Australian Square Kilometre Array Pathfinder Evolutionary Map of the Universe survey. We derived the spectral index from the matched sources in the GAMA-23 field for the first time. We have studied the statistical properties of diffuse Galactic synchrotron emission (DGSE) in the GAMA-23 field using six different latitude target pointings. We fitted DGSE in the form of power law $C_{\ell} = A(1000/{\ell})^{{\beta}}$. Our DGSE amplitude range of $A = 1-20\,{\text{mK}}^2$ and $\beta$ varies between $1.5$ and $2.4$. This work presents one of the first characterizations of DGSE in the southern sky at 325\,MHz, setting a precedent for foreground modeling required for sensitive radio observations with the upcoming SKA. 
\end{abstract}

\keywords{methods: observational -- data analysis techniques: interferometric -- radio continuum: galaxies -- catalogs -- (cosmology:) diffuse radiation}

\section{Introduction} \label{sec:intro}
Radio surveys provide a dust-unbiased view of both Galactic and extragalactic skies. At sub-GHz to a few GHz frequencies, emission is primarily governed by synchrotron radiation from cosmic-ray electrons  \citep{1999A&A...345..380S}, tracing two key populations: active galactic nuclei (AGN) and star-forming galaxies \citep[SFGs;][]{2007MNRAS.375..931M, 2016MNRAS.457..730P}. In AGN, this emission originates from relativistic jets and is crucial to understand feedback processes in galaxy evolution \citep{2006MNRAS.370..645B, 2012ARA&A..50..455F, 2014MNRAS.445..955B, 2017NatAs...1E.165H}. In SFGs, synchrotron radiation from supernova-accelerated electrons serves as a reliable tracer of star-formation rates \citep{2009MNRAS.397.1101G, 2018MNRAS.475.3010G, 2021A&A...648A...6S}, unaffected by dust extinction. Supernova remnants \citep{2010MmSAI..81..374W, 2017A&A...605A..58A},  diffuse Galactic synchrotron \citep{2005ApJ...625..575S}, and transient sources \citep{2014MNRAS.438..352B} are the cause of radio emission within the Galaxy. These surveys are also essential to construct accurate radio sky models and to identify transient or variable sources across repeated epochs \citep{2011ApJ...742...49T}. Some of the telescopes used to conduct surveys are the upgraded Giant Meterwave Radio Telescope \citep[uGMRT;][]{2017CSci..113..707G}, Very Large Array \citep[VLA;][]{2011ApJ...739L...1P}, LOw Frequency Array \citep[LOFAR;][]{2013A&A...556A...2V}, Australian Square Kilometre Array Pathfinder \citep[ASKAP;][]{2008ExA....22..151J, 2014PASA...31...41H, hotan2021australian}, the Murchison Wide-field Array \citep[MWA;][]{2013PASA...30...31B}, Australia Telescope Compact Array \citep[ATCA;][]{2011MNRAS.416..832W}, MeerKAT \citep{2016mks..confE...1J}, and upcoming Square Kilometre Array \citep[SKA;][]{2020PASA...37....2W}. \\
As we prepare for the upcoming SKA \citep{2015aska.confE.171C, 2015aska.confE...5C}, its precursors and pathfinders are pushing the frontiers through deep and wide-field radio surveys \citep{2013PASA...30...20N}. The current generation of telescopes cover surveys including the 150\,MHz LOFAR Two-meter Sky Survey \citep[LoTSS;][]{2019A&A...622A...1S, 2022A&A...659A...1S} with LoTSS deep field \citep{2025MNRAS.542.2245A, 2025A&A...695A..80S}, MeerKAT MIGHTEE continuum survey \citep{2025MNRAS.536.2187H} and deep polarized sky \citep{2024MNRAS.528.2511T}, TIFR GMRT Sky Survey (TGSS) 150\,MHz All-Sky radio survey \citep{2017A&A...598A..78I} from uGMRT, GaLactic and Extragalactic All-sky Murchison Widefield Array (GLEAM) survey \citep{2017MNRAS.464.1146H, 2022PASA...39...35H, 2025PASA...42..158D} at 72–231\,MHz and at 300\,MHz from MWA, the Rapid ASKAP Continuum Survey \citep[RACS;][]{2020PASA...37...48M, 2023PASA...40...34D}, the 843\,MHz Sydney University Molonglo Sky Survey \citep{2003MNRAS.342.1117M}, the 1.4 GHz NRAO VLA Sky Survey and \citep[NVSS;][]{1998AJ....115.1693C}. Despite the wealth of data from these surveys, challenges persist. There is substantial sky overlap between multiple northern and southern radio surveys. However, overlapping regions are observed at low elevation pointings or near the celestial equator, resulting in low-quality data. In addition, there are relatively few deep field observations in the overlapping regions. This limits comprehensive multifrequency studies over common regions of the sky. \\
The Galaxy and Mass Assembly Project\footnote{\url{http://www.gama-survey.org/}} \citep[GAMA;][]{2016MNRAS.455.3911D} is a comprehensive multiband imaging and spectroscopic survey targeting different fields in the Southern sky, covering ${\sim}286\,\text{deg}^2$ area. One of the popular fields is the GAMA-23 field due to the richness in multiwavelength data. Most of the recent surveys from ASKAP cover the Southern sky; this includes the Evolutionary Map of the Universe \citep[EMU;][]{2021PASA...38...46N, 2025PASA...42...71H} survey, the RACS \citep{2021PASA...38...58H} and MeerKLASS L-band Continuum survey \citep{2025arXiv251217685M}. 
Our work observed the GAMA-23 field at 325\,MHz with the uGMRT. The observed area overlaps with the ASKAP GAMA-23 field survey \citep{2022MNRAS.512.6104G}. This paper will complement available surveys of the Southern sky, such as TGSS at 150\,MHz \citep{2017A&A...598A..78I} and ASKAP GAMA-23 field survey at 887.5\,MHz \citep{2022MNRAS.512.6104G}.\\ 
Nonideal radio sky models are not the only challenge; foregrounds are also among the dominant barriers to detecting the 21 cm signal from the Epoch of Reionization. Foregrounds are of two types -- Galactic and extragalactic, with Galactic foregrounds, namely diffuse Galactic synchrotron emission (DGSE), dominating by 70\% \citep{2008MNRAS.389.1319J}. The DGSE is characterized by fluctuations in the magnetic field and the cosmic-ray electron density in our Galaxy \citep{1992ARA&A..30..575C, 2013A&A...558A..72I}. In 21 cm experiments, mapping the spatial and spectral characteristics of DGSE is crucial for modeling the foregrounds and mitigating them to detect the 21 cm signal \citep{di200421, 2009ApJ...695..183B, 2012ApJ...752..137M, 2016MNRAS.458.1057O, Chapman2020Foregrounds}. The foregrounds dominate by 4-5 orders of magnitude and are smooth, following a power law \citep{2002ApJ...564..576D, 2008MNRAS.385.2166A}. However, previous studies have characterized the spatial properties of DGSE using the angular power spectrum (APS) from uGMRT observations \citep{2019MNRAS.487.4102C, 2020MNRAS.494.1936C, 2020MNRAS.495.4071M, 2025MNRAS.544.3617S}. In particular, \citep{2019MNRAS.490..243C} studied frequency evolution of the DGSE APS amplitude and identified a spectral break at ${\nu}_\text{break}=405$\,MHz in the ELAIS-N1 field from uGMRT Band-3 wideband observations. Motivated by these studies, the present work focuses on the spatial properties of DGSE in the GAMA-23 field. We fitted a power law ($C_{\ell} = A(1000/{\ell})^{{\beta}}$) and investigated whether the spatial characteristics are consistent with those estimated in other high Galactic latitudes.\\% \citep{2012MNRAS.426.3295G}.\\
This paper provides the mosaic, source catalog and spatial feature of DGSE in the GAMA-23 field using uGMRT data at Band-3. Section~\ref{sec:observation} offers a detailed overview of the uGMRT observations in Band-3 from this work. Data reduction and imaging have been addressed in Section~\ref{sec:calibration}, while Section~\ref{sec:catalog} exhibits the extracted catalog from the mosaic image and source classification. Section~\ref{sec:comparison} discusses the comparison of this catalog with other available catalogs and surveys, verifying the accuracy of flux density and position of sources. Section~\ref{sec:count} calculates the completeness of the catalog and presents corrected source counts. Section~\ref{sec:DGSE_APS} describes the characterization of DGSE in terms of power law at angular scale.  Section~\ref{sec:conclusion} provides the conclusion of this work.  

\section{Observation} \label{sec:observation}
%%%%%%%%%%%%%%%%%%%%%%%%%%%%%%%%%%%%%%%%%%%%%%%%%%%%%%%%%%%%%%%%%%%%%%%%%%%%%%%%
\begin{figure*}
    \centering
    \includegraphics[width=5.8in]{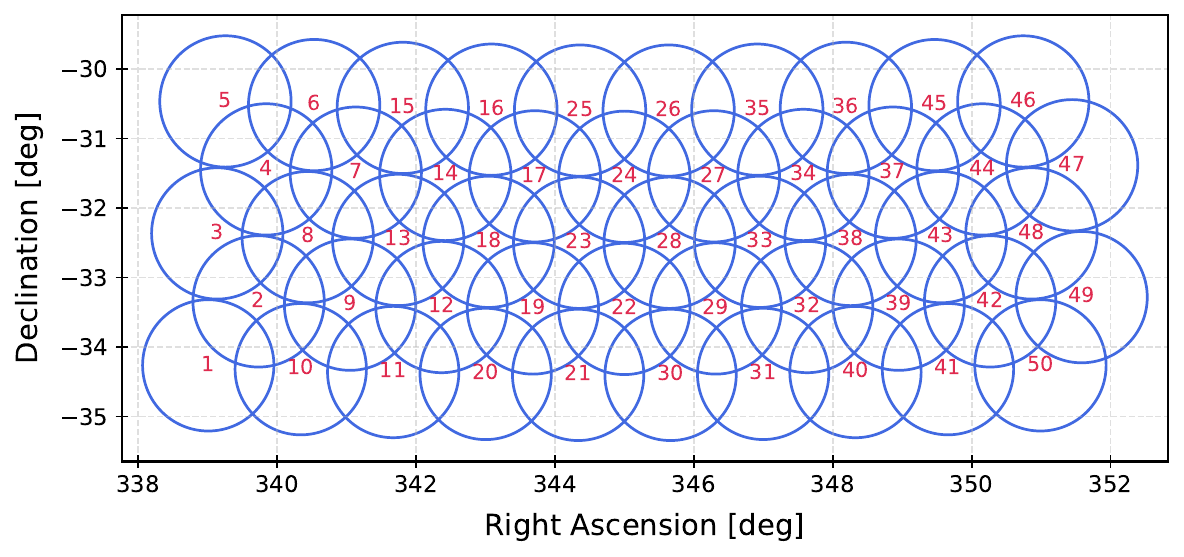}
    \caption{All 50 target pointings in the GAMA-23 field are shown as blue circles with their corresponding field IDs in red. This observation consists of ${\sim}\,56$ minutes per pointing using uGMRT Band\,-\,3 in observation cycle 32.}
    \label{fig:Fig1}
\end{figure*}
%%%%%%%%%%%%%%%%%%%%%%%%%%%%%%%%%%%%%%%%%%%%%%%%%%%%%%%%%%%%%%%%%%%%%%%%%
This paper used 33 hr of archival data of the GAMA-23 field ($\alpha_{J2000} = 345^{\circ}$, $\delta_{J2000} = -32.50^{\circ}$) observed with the uGMRT in Band-3 \citep{2017CSci..113..707G}. The observations were taken during cycle 32 (Proposal Code – 32\_060\footnote{\url{https://naps.ncra.tifr.res.in/goa/}}) with the GMRT software backend \citep[GSB;][]{2010ExA....28...25R} at 325 MHz central frequency with a bandwidth of 32 MHz per pointing. Table~\ref{table:tab1} describes the information on uGMRT observation parameters and position of calibrators in degree units. The observations have been taken over 6 nights between 2017 July 28 and August 27, involving 50 different pointing centers. For each pointing, the total on-source time is almost $56$ minutes. Due to the short integration time per pointing and the frequency-dependent variations in the primary beam, the wideband data showed residual spectral artifacts in individual pointings. Thus, we limited our analysis to the GSB data to mitigate these effects, as the narrower bandwidth reduces calibration complexity, including gain and spectral features. The final image mosaicked with 50 target pointings at range $337.85^{\circ}$ $\leq \alpha_\text{J2000} \leq$ $352.72^{\circ}$ and -35.34$^\circ$ $\leq \delta_\text{J2000} \leq$ -29.52$^\circ$. 
Figure ~\ref{fig:Fig1} presents all of the 50 pointings with blue circles and each pointing numbered with red. For the flux calibrator, either $3\text{C}\,48$ or $3\text{C}\,468.1$, or both were observed in the beginning and the end of each observation. The phase calibrator $0025-260$, around the target pointings, is observed for 5 minutes between three or four target pointings of the GAMA-23 field. 
%%%%%%%%%%%%%%%%%%%%%%%%%%%%%%%%%%%%%%%%%%%%%%%%%%%%%%%%%%%%%%%%%%%%%%%%%%%%%%%%%%%%%%%%%%%%%%
\begin{table}
\caption{uGMRT observation parameters.}
\centering
\begin{tabular}{ll}
\hline\hline
Project code & 32\_060 \\ [0.3ex]
Observation date & 2017 Jul 28 \\ 
                 & 2017 Aug 02, 14, 15, 19, 27 \\ [0.3ex]
Total observation time & 33 hr \\ [0.3ex]
Central frequency & $325$\,MHz \\ [0.3ex] %$306-339$\,MHz\\ [0.3ex]
Channels & 256 \\ [0.3ex] %256
Mosaic field of view & $65.31\,\text{deg}^{2}$ \\ [0.3ex] %$75.84\,\text{deg}^{2}$ \\
%\hline
Flux calibrator & 3C48 and 3C468.1 \\ 
                 & $24.42^{\circ} +33.16^\circ$ \\%($01^h 37^m 41^s + 33^\circ 09' 35''$) \\ %$24.42^{\circ} +33.16^\circ$ \\%
                 & $357.73^{\circ} +64.67^\circ$ \\ [0.3ex] %($23^h 50^m 55^s + 64^\circ 40' 18''$)\\ [0.3ex]
Phase calibrator & $0025-260$ \\
                 & $6.45^{\circ} -26.04^\circ$ \\ [0.3ex] %($0^h 25^m 49^s - 26^\circ 02' 13''$) \\ [0.3ex]
Total pointing center & 50 \\
\hline\hline
\end{tabular}
\label{table:tab1}
\end{table}

\section{Data reduction and Imaging} \label{sec:calibration}
%%%%%%%%%%%%%%%%%%%%%%%%%%%%%%%%%%%%%%%%%%%%%%%%%%%%%%%%%%%%%%%%%%%%%%%%%%%%%%%%%%%%%%%%%%%%%%%%%
\begin{figure*}
\centering

\begin{tikzpicture}

\node[anchor=south west,inner sep=0] (image)
at (0,0)
{\includegraphics[width=\linewidth]{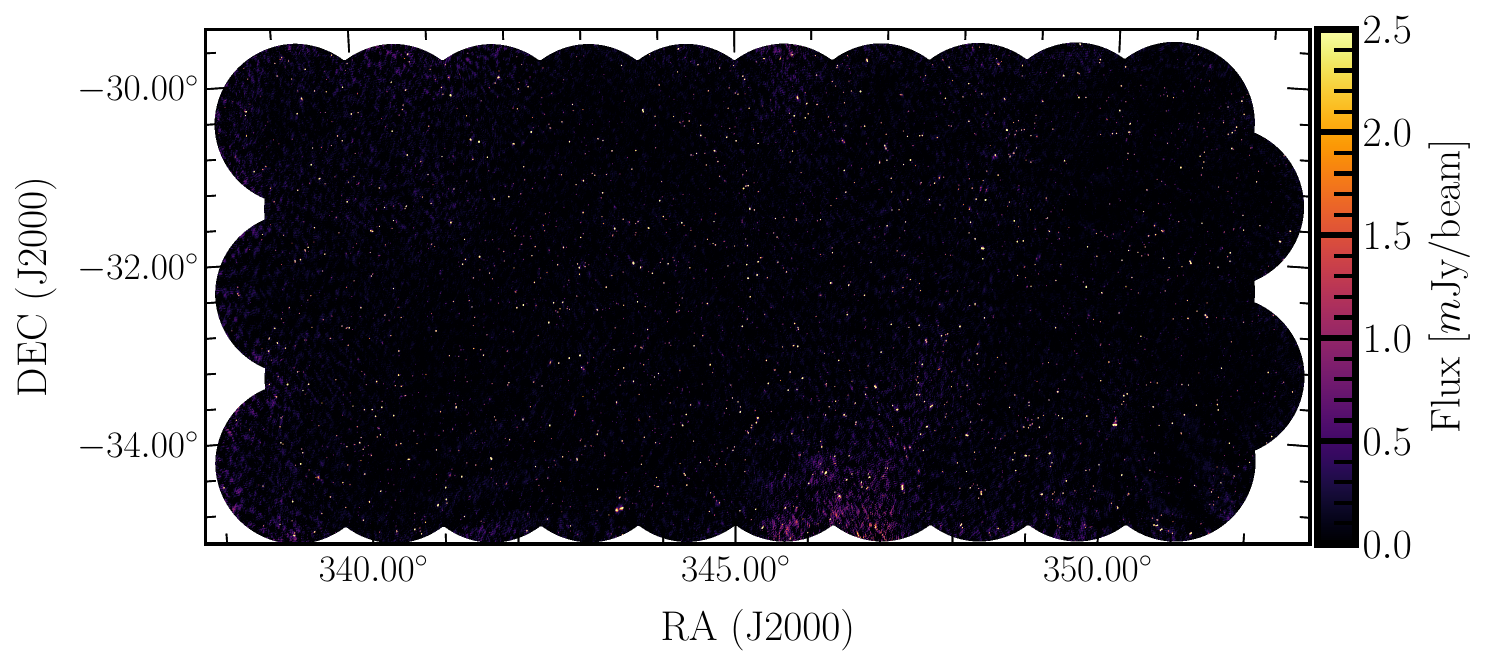}};

%%%%%%%%%%%%%%%%%%%%%%%%%%%%%%%%%%%%%%%%%%%%%%%%%%
% SOURCE BOXES (ALL SAME SIZE)
%%%%%%%%%%%%%%%%%%%%%%%%%%%%%%%%%%%%%%%%%%%%%%%%%%

% Source 1
\draw[cyan, very thick]
(3.40,4.70) rectangle ++(0.40,0.40);

% Source 2
\draw[cyan, very thick]
(4.35,5.10) rectangle ++(0.40,0.40);

% Source 3
\draw[cyan, very thick]
%(10.60,3.40) rectangle ++(0.40,0.40);
(7.35,1.73) rectangle ++(0.40,0.40);

% Source 4
\draw[cyan, very thick]
(14.30,4.70) rectangle ++(0.40,0.40);

% Source 5
\draw[cyan, very thick]
(6.43,3.50) rectangle ++(0.40,0.40);

%%%%%%%%%%%%%%%%%%%%%%%%%%%%%%%%%%%%%%%%%%%%%%%%%%
% CONNECTOR LINES (ALL SAME STYLE)
%%%%%%%%%%%%%%%%%%%%%%%%%%%%%%%%%%%%%%%%%%%%%%%%%%
% Source 1
\draw[cyan, thick]
(3.60,4.70) -- (1.7,-0.6);

% Source 2
\draw[cyan, thick]
(4.55,5.10) -- (5.4,-0.6);

% Source 3
\draw[cyan, thick]
%(10.80,3.40) -- (9.0,-1.0);
(7.55,1.73) -- (8.4,-0.6);

% Source 4
\draw[cyan, thick]
(14.55,4.70) -- (12.6,-0.6);

% Source 5
\draw[cyan, thick]
(6.63,3.50) -- (16.2,-0.6);
\end{tikzpicture}
%\vspace{0.1cm}

%%%%%%%%%%%%%%%%%%%%%%%%%%%%%%%%%%%%%%%%%%%%%%%%%%
% MIDDLE PANEL : SOURCE CUTOUTS
%%%%%%%%%%%%%%%%%%%%%%%%%%%%%%%%%%%%%%%%%%%%%%%%%%
\begin{minipage}{0.17\linewidth}
\centering
\includegraphics[width=\linewidth]{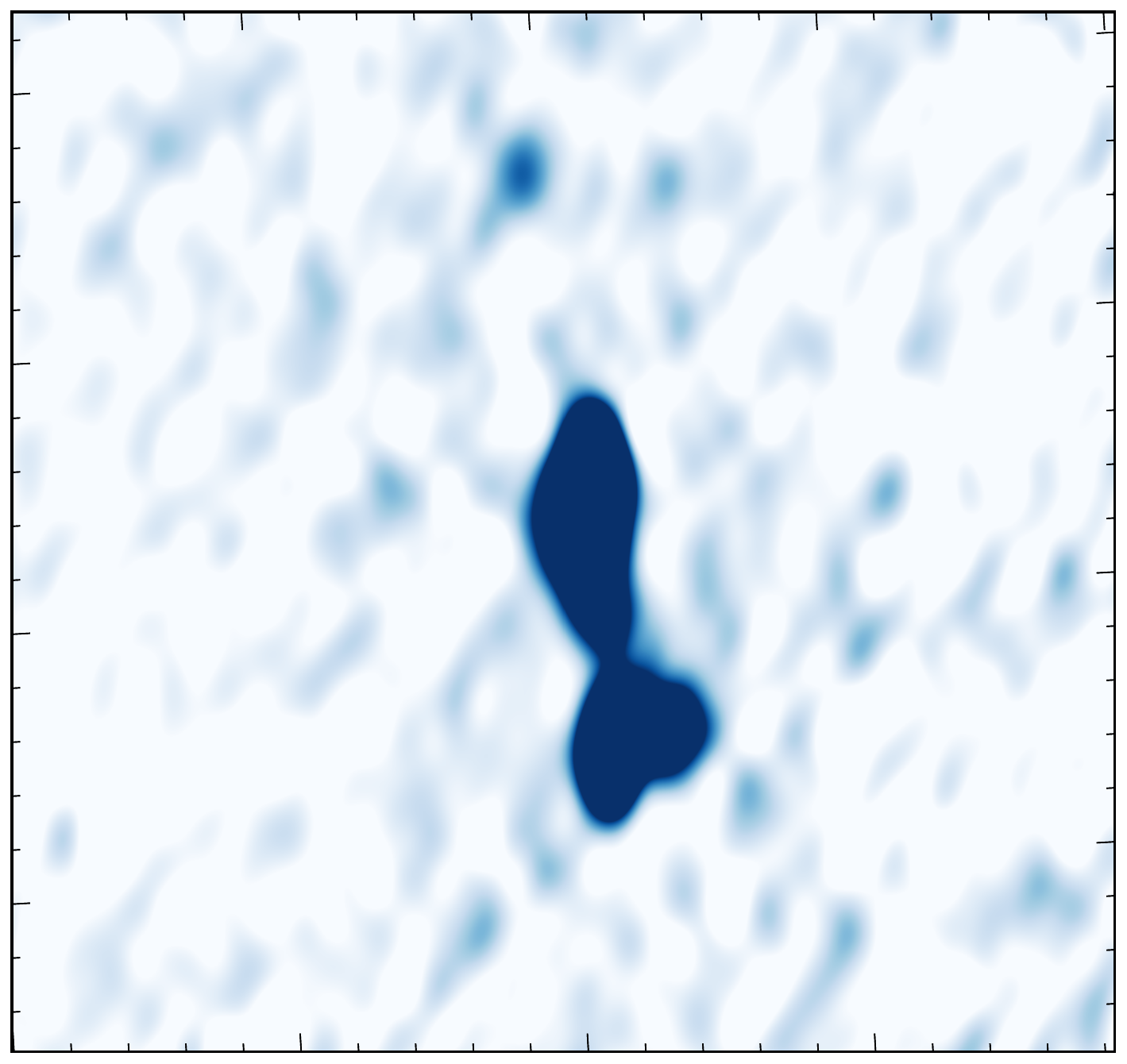}
\end{minipage}
\hfill
\begin{minipage}{0.17\linewidth}
\centering
\includegraphics[width=\linewidth]{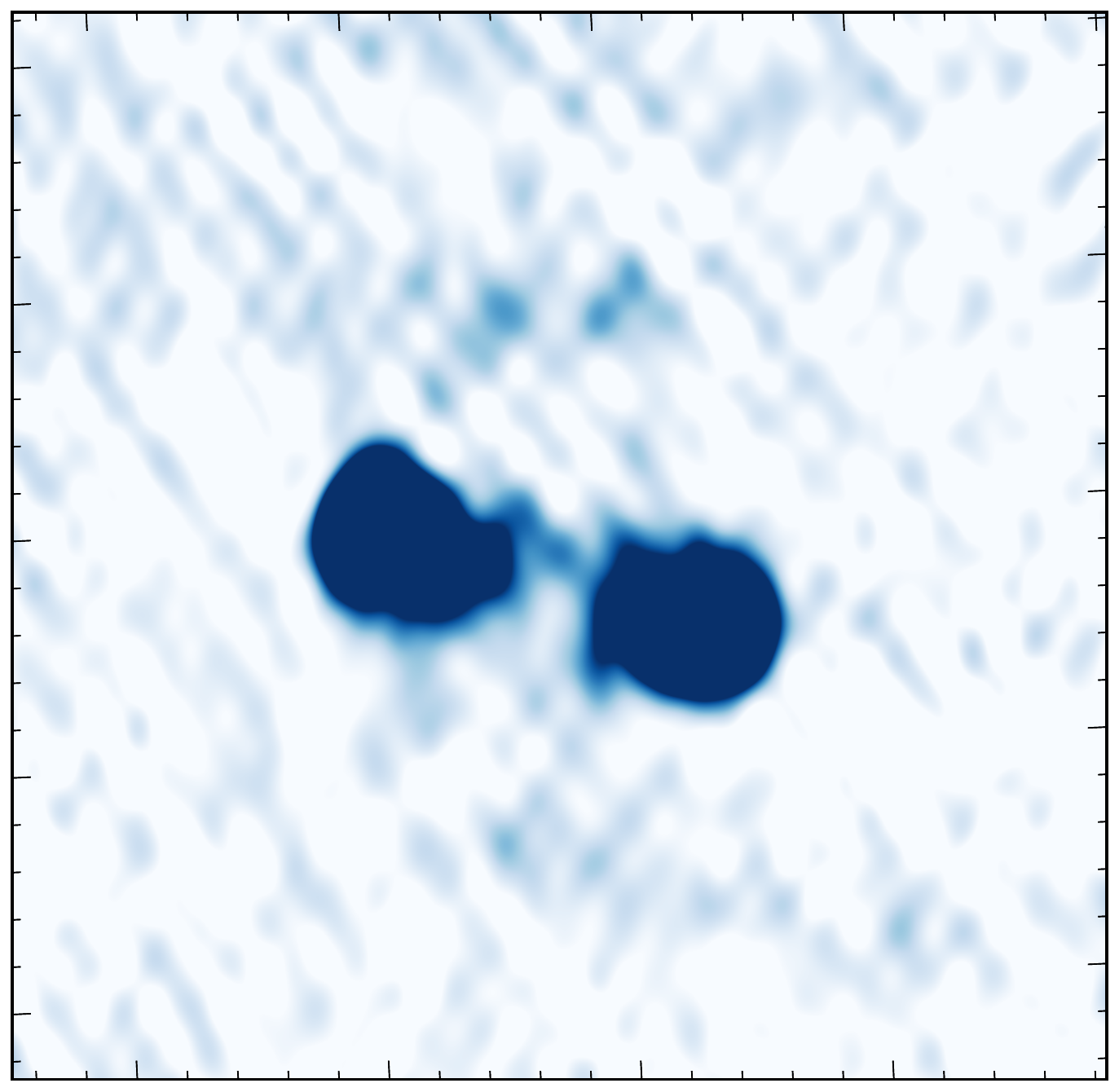}
\end{minipage}
\hfill
\begin{minipage}{0.17\linewidth}
\centering
\includegraphics[width=\linewidth]{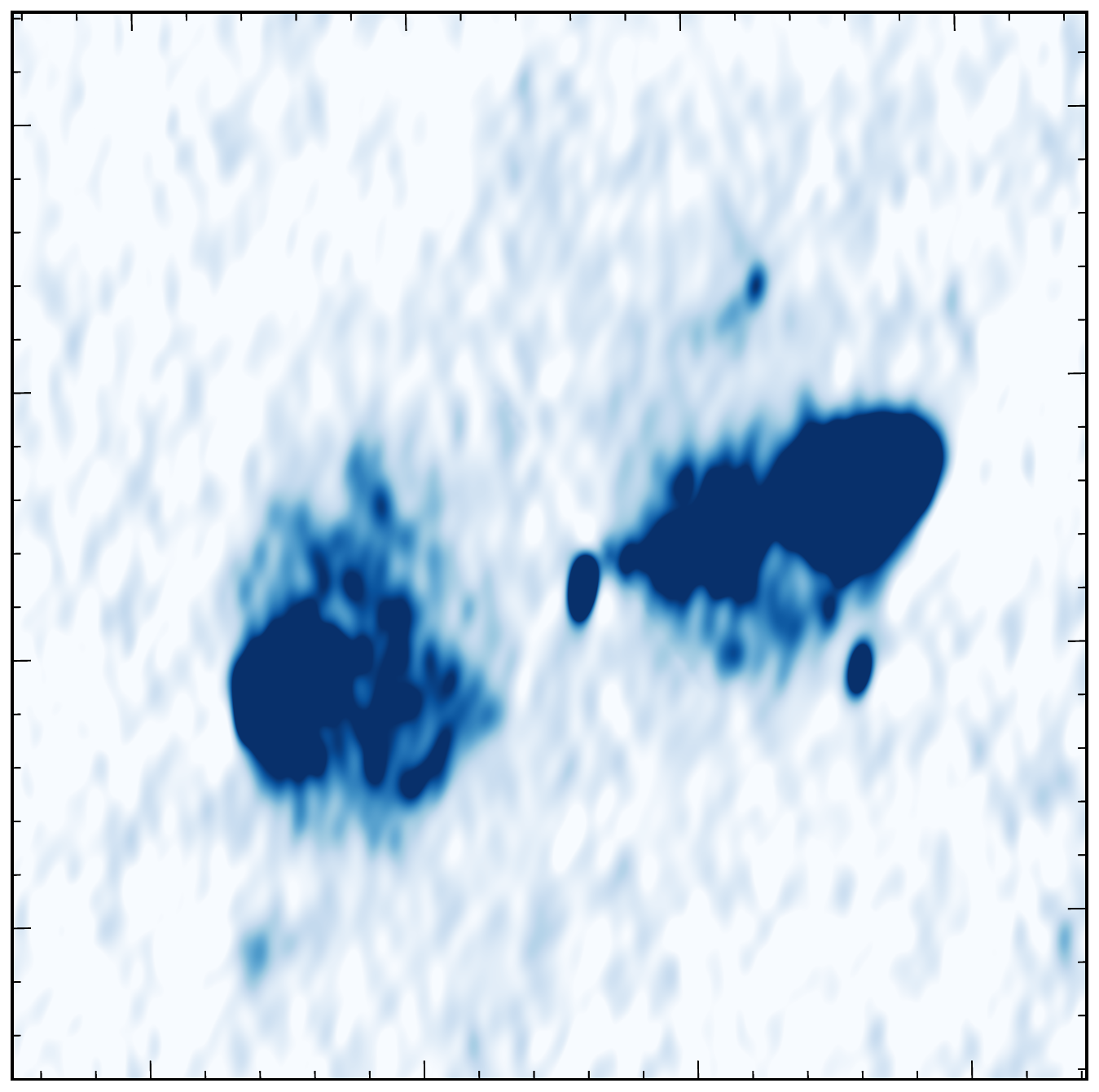}
\end{minipage}
\hfill
\begin{minipage}{0.17\linewidth}
\centering
\includegraphics[width=\linewidth]{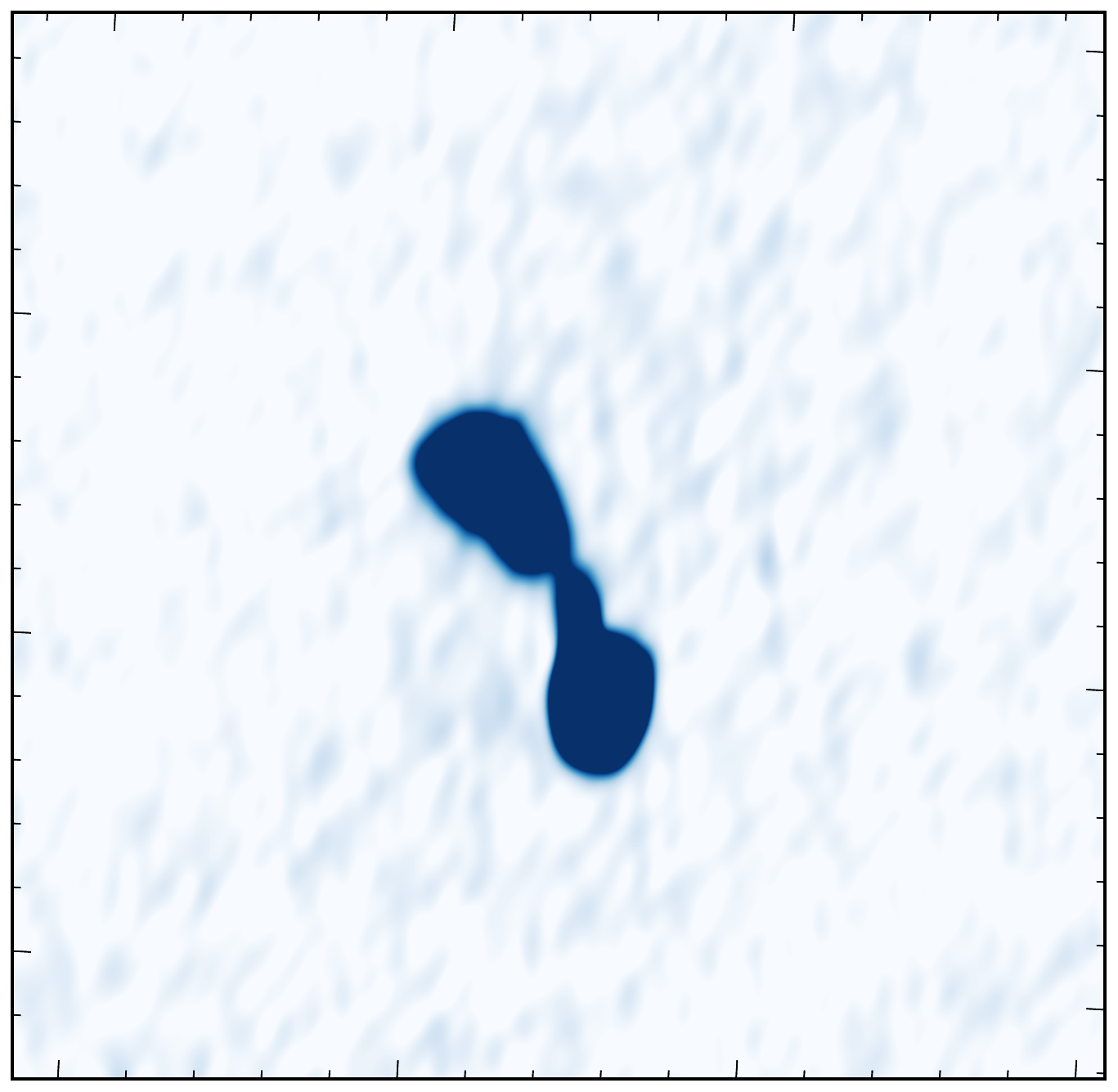}
\end{minipage}
\hfill
\begin{minipage}{0.17\linewidth}
\centering
\includegraphics[width=\linewidth]{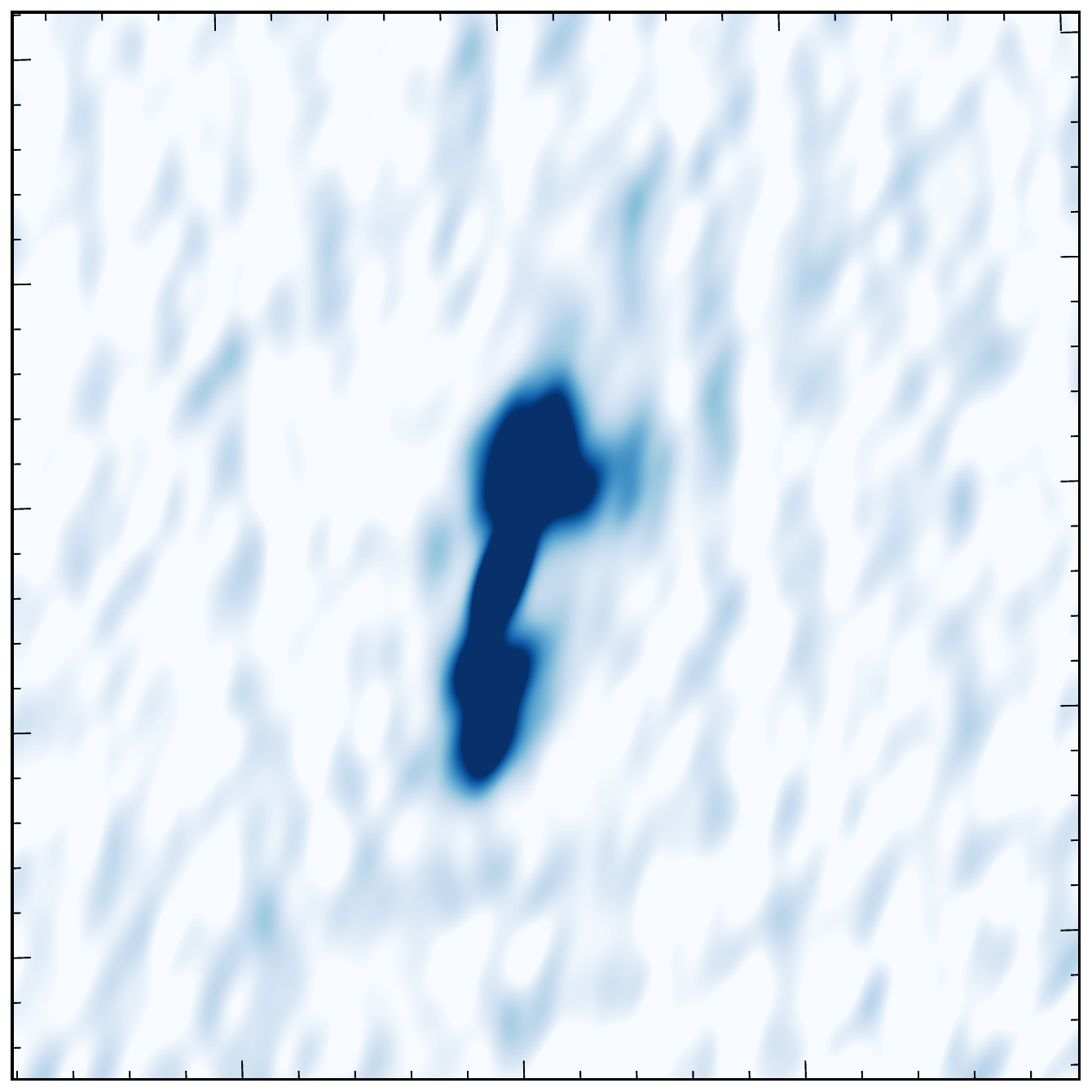}
\end{minipage}
\vspace{0.3cm}
%%%%%%%%%%%%%%%%%%%%%%%%%%%%%%%%%%%%%%%%%%%%%%%%%%
% BOTTOM PANEL : RMS IMAGE
%%%%%%%%%%%%%%%%%%%%%%%%%%%%%%%%%%%%%%%%%%%%%%%%%%
\includegraphics[width=\linewidth]{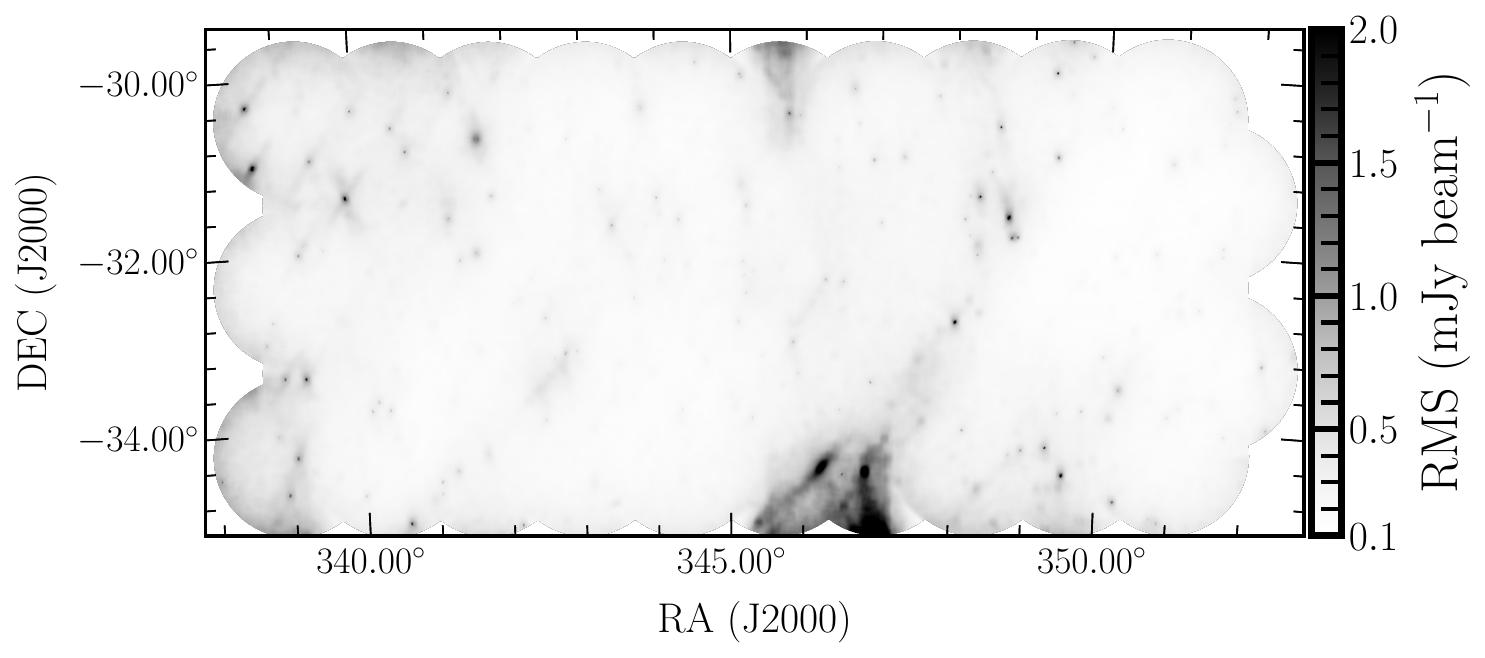}
\caption{Upper Panel: The mosaic image of GAMA-23 field consists of 50 pointings produced after DD-calibration at 325\,MHz. The central off-source RMS noise attained is 109\,${\mu}$Jy$\,\mathrm{beam}^{-1}$ with area coverage of $65.31\,\text{deg}^2$. The FoV of the mosaic image is $14.9{^\circ}{\times}5.8{^\circ}$ and the restoring beam size is $15.8'' {\times} 8.3''$. Middle Panel: Some sample radio sources identified in the GAMA-23 field. Lower Panel: The background RMS noise map of the mosaic image. The high RMS noise has been observed around the bright sources and toward the edges of FoV.}
\label{fig:Fig2}
\end{figure*}
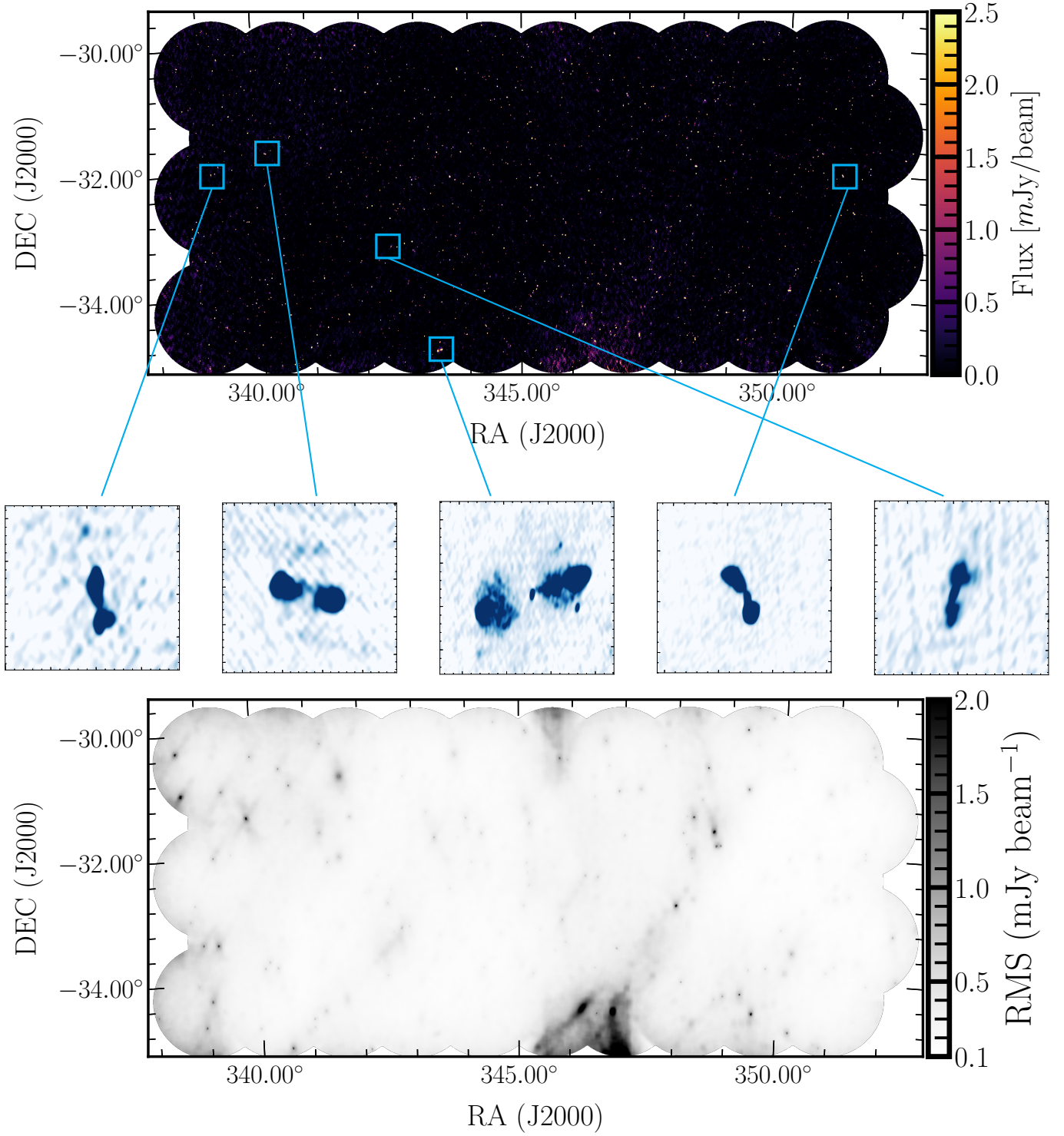
%%%%%%%%%%%%%%%%%%%%%%%%%%%%%%%%%%%%%%%%%%%%%%%%%%%%%%%%%%%%%%%%%%%%%%%%%%%%%%%%%%%%%%%%%%%%%%%%%%%
We used source peeling and atmospheric modeling \citep[{\tt\string SPAM};][]{2009A&A...501.1185I} to calibrate our data. SPAM is {\tt\string AIPS} \citep{2003ASSL..285..109G} based Python scripted semi-automated pipeline. SPAM considers the ionospheric delay correction, which is important to reduce uncertainties in pointing errors and phase variations across the field of view (FoV). The SPAM assumption is that the ionosphere is a 2D refractive screen. Therefore, provide solutions for both ionospheric phase delay and instrument gain. Hence, it works as a direction-dependent (DD) calibration pipeline \citep{2014ASInC..13..469I}.  \\
For each pointing, data undergoes the calibration procedure. The bandwidth of GSB data for each pointing is 32\,MHz. In the preprocessing stage, the data were examined for bad antennas, radio frequency interference, and corrupted data in time intervals. The data were then averaged in the time and frequency domains, resulting in 42 channels with a channel width of 781.25\,kHz \citep[similar to][]{2019MNRAS.487.4102C, 2020MNRAS.495.4071M}. Then the calibrator and target pointing have been separated. In the initial calibration, the best scan of the primary calibrator (i.e., 3C48) has been set to the flux density scale using the flux-scale model \cite{2012MNRAS.423L..30S}.
Furthermore, the gain and bandpass solutions have been derived and implemented on the target pointing. In the main pipeline, the target pointing was iteratively imaged as a part of the self-calibration procedure to further improve the calibration and image quality. Following this, multiple bright sources across each pointing are used to model and correct for phase distortion due to the ionosphere. In the final step, the primary-beam corrected image and calibrated visibilities are produced. This process has been applied to each target pointing. \citep[same process as followed in][]{2019MNRAS.487.4102C, 2020MNRAS.495.4071M} \\
Following the DD-calibration performed with SPAM, we used WsClean\footnote{\url{https://gitlab.com/aroffringa/wsclean}} \citep{2014MNRAS.444..606O} for imaging of calibrated visibilities of each target pointing. We used the multiscale and multifrequency deconvolution algorithm implemented in WsClean \citep{2017MNRAS.471..301O}. We then implemented primary-beam correction to the image of each target pointing. We used a 20\% cut-off for the primary-beam response. We create the final image by combining each target pointing together into a mosaic \citep[similarly in][]{2023MNRAS.525.5311S}. Of the 50 pointing images, only 30 and 31 were highly artifact affected but usable. To ensure uniform weighting across the mosaic image, we weighted each primary-beam corrected image by noise variance, calculated as the square of the primary-beam response. Upper panel of Figure~\ref{fig:Fig2} presents the mosaic image after combining the primary-beam corrected images of each target pointing. The minimum central off-source RMS noise achieved is 109\,$\mu$Jy beam$^{-1}$. The FoV of the uGMRT mosaic image at 325\,MHz is $14.9{^\circ}{\times}5.8{^\circ}$. The restoring beam of the final mosaic is $15.8'' \times 8.3''$, and the position angle is $-0.85^\circ$. Each individual pointing covers FoV of $1.89\,\text{deg}$ with an area $2.81\,\text{deg}^2$ while the mosaic image covers the area of $65.31\,\text{deg}^2$. The middle panel of Figure~\ref{fig:Fig2} shows sample radio sources detected in the field.

\section{Source catalog} \label{sec:catalog}
%%%%%%%%%%%%%%%%%%%%%%%%%%%%%%%%%%%%%%%%%%%%%%%%%%%%%%%%%%%%%%%%%%%%%%%%%%%%%%%%%%%%%%%%%%

\begin{deluxetable}{ccccccccccc}
%\digitalasset
\tablewidth{0pt} % Allow table to expand to natural width
\tabletypesize{\scriptsize} % Reduce font size to fit more horizontally
\tablecaption{Sample source catalog from mosaic image of GAMA-23 field with uGMRT in Band-3}
\tablehead{
\colhead{Source Id} & \colhead{R.A.} & \colhead{Decl.} & \colhead{Total\_Flux} & \colhead{Peak\_Flux} & \colhead{\text{Maj}} & \colhead{\text{Min}} & \colhead{\text{PA}} & \colhead{Isl\_rms} \\ 
\colhead{} & \colhead{(deg $\pm$ arcsec)} & \colhead{(deg $\pm$ arcsec)} & \colhead{(mJy)} & \colhead{(mJy beam$^{-1}$)} & \colhead{(arcsec)} & \colhead{(arcsec)} & \colhead{(deg)} & \colhead{(mJy beam$^{-1}$)}
}
\startdata
0 & 338.1126 $\pm$ 0.26 & -32.1327 $\pm$ 0.64 & 7.26 & 7.29 & 14.58 & 8.79 &  167.60 & 0.62 \\
1 & 338.2543 $\pm$ 0.52 & -30.5169 $\pm$ 0.56 & 16.30 & 9.38 & 15.40 & 14.50 &  15.66 & 0.74\\
2 & 338.1585 $\pm$ 0.62 & -32.5315 $\pm$ 1.32 & 2.45 & 2.59 & 13.89 & 8.75 &  162.17 & 0.50 \\
\enddata
\tablecomments{The extracted source catalog includes the columns of source IDs, source positions, integrated and peak flux densities, source sizes (Min and Maj axes), position angle (PA), and corresponding RMS noise of fitted islands. The full catalog will be fetchable in table format with the online version of this paper.}
\label{table:tab2}
\end{deluxetable}

We used PyBDSF\footnote{\url{https://www.astron.nl/citt/pybdsf/}} \citep{2015ascl.soft02007M}, the source extractor package, to generate a source catalog from the mosaic image of the GAMA-23 field. The sliding box of size 180 pixels every 50 pixels $\tt{rms\_box = (180,50)}$ was used to compute the local background RMS noise variation. To prevent artifacts from being mistaken for real sources, we used a small box (${\tt rms\_box\_bright = (35,7)}$) around the bright sources. To suppress artifacts and spurious sources detection around bright sources, we applied an adaptive threshold of 150$\sigma$. This increases the detection threshold locally for pixels exceeding a specific signal-to-noise ratio (SNR). PyBDSF detects contiguous regions of emission (islands) above the defined pixel threshold and models each island using multiple Gaussian components. Islands of emission were detected at a 3${\sigma}_\text{rms}$ threshold, while Gaussian components were fitted to peaks above 5${\sigma}_\text{rms}$. To account for the varying PSF, we set ``$\tt{psf\_vary\_do = True}$'' to handle beam variations across the FoV \citep[parameters equivalent to][]{2025MNRAS.544.3617S}.\\
An RMS map was generated to depict the spatial variation of background RMS noise over the field. Lower panel of Figure~\ref{fig:Fig2} shows that the RMS increases near bright sources and toward the field edges. The higher noise around bright sources is mainly due to imaging artifacts. Beyond the contributions from resolution and thermal noise, the precision of the extracted source catalog is also limited by confusion noise, arising from the background sky brightness fluctuations that occur due to multiple faint, unresolved sources within the synthesized beam. This can be calculated as described in \citep{2012ApJ...758...23C}:
%%%%%%%%%%%%%%%%%%%%%%%%%%%%%%%%%%%%%%%%%%%%%%%%%%%%%%%%%%%%%%%%%%%%%%%%%%%%%%%%%%%%%%%%%%%%%%%%%%%%%%
\begin{equation}
\label{eqn:2}
\begin{aligned}
\sigma_{CN} = 1.2\left(\frac{\nu}{3.02\,\text{GHz}}\right)^{-0.7} \left(\frac{\theta}{8\,\text{arcsec}}\right)^{10/3}\,\mu\text{Jy/beam}
\end{aligned}
\end{equation}
%%%%%%%%%%%%%%%%%%%%%%%%%%%%%%%%%%%%%%%%%%%%%%%%%%%%%%%%%%%%%%%%%%%%%%%%%
where $\nu$ is the observational frequency, and $\theta$ is the FWHM of the beam of telescope. It yields a confusion noise ${\sim}\,55\,{\mu}$Jy beam$^{-1}$. Thus, confusion noise is lower than the central off-source RMS noise estimated in this paper (i.e., 109\,$\mu$Jy beam$^{-1}$). Thus, the results from this work are above the sensitivity limit imposed by confusion noise. From the uGMRT Band-3 mosaic image, we have extracted a catalog containing 5741 sources in the GAMA-23 field, which covers an area $65.31\,\text{deg}^2$ with a detection threshold of ${\geq}$\,5$\sigma_\text{rms}$. A sample of the source catalog is presented in Table~\ref{table:tab2} (the complete catalog is available in the electronic version of this paper).

\subsection{Compact vs Resolved sources} \label{sec:classification}
For compact sources, the ratio of integrated-to-peak flux density is anticipated to be unity. The observed spread around unity is due to parameter estimation errors, calibration errors and noise fluctuations. In addition, residual ionospheric smearing can reduce the peak flux density, contributing to the observed scatter in the flux density ratio. In this section, we studied the flux density ratio of integrated-to-peak flux (${S_{\text{int}}}/{S_{\text{peak}}}$) of extracted catalog sources as a function of (${S_{\text{peak}}}/{\sigma_L}$), where ${\sigma_L}$ stands for RMS noise. We investigated the integrated-to-peak flux density ratio of high-signal-to-noise compact sources across the GAMA-23 field mosaic image. We used the method discussed in \cite{2015MNRAS.453.4020F, 2019PASA...36....4F} to detect resolved sources at 3${\sigma}_R$. Above this threshold, all sources are considered compact. The equation is given as follows:
%%%%%%%%%%%%%%%%%%%%%%%%%%%%%%%%%%%%%%%%%%%%%%%%%%%%%%%%%%%%%%%%%%%%%%%%%%%%%%%%%%%%%%%%%%%%%%%%%%%%
\begin{equation}
\begin{aligned}
\label{eqn:3}
\sigma_R &= \sqrt{\left(\frac{\sigma_{S_{\text{int}}}}{S_{\text{int}}}\right)^2 + \left(\frac{\sigma_{S_{\text{peak}}}}{S_{\text{peak}}}\right)^2}
\end{aligned}
\end{equation}
%%%%%%%%%%%%%%%%%%%%%%%%%%%%%%%%%%%%%%%%%%%%%%%%%%%%%%%%%%%%%%%%%%%%%%%%%%%%%%%%%%%%%%%%%%%%%%%%%%%%
%%%%%%%%%%%%%%%%%%%%%%%%%%%%%%%%%%%%%%%%%%%%%%%%%%%%%%%%%%%%%%%%%%%%%%%%%%%%%%%%%%%%%%%%%%%%%%%%%%%%
\begin{figure}
    \centering
    \includegraphics[width=3.35in]{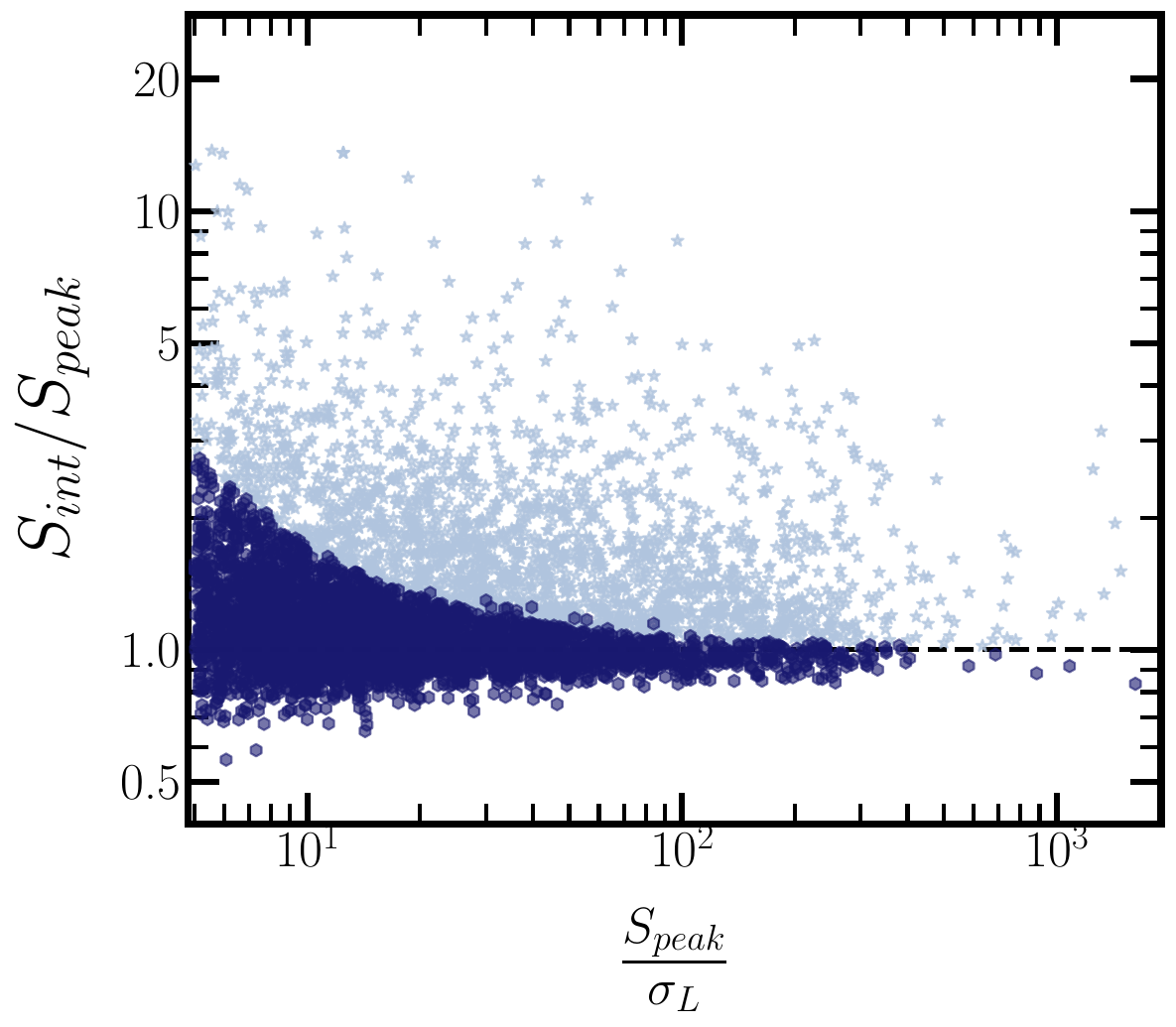}
    \caption{The ratio of integrated-to-peak flux density ($S_{\mathrm{int}}/S_{\mathrm{peak}}$) is shown as a function of SNR ratio ($S_{\mathrm{peak}}/\sigma_L$) for sources in our catalog. Resolved sources are found to be 1966 (steel-blue stars), while compact sources are 3775 in number (navy hexagons).}
    \label{fig:Fig5}
\end{figure}
%%%%%%%%%%%%%%%%%%%%%%%%%%%%%%%%%%%%%%%%%%%%%%%%%%%%%%%%%%%%%%%%%%%%%%%%%%%%
where $\sigma_{S_{\text{int}}}$ and $\sigma_{S_{\text{peak}}}$ represent uncertainties in the integrated flux values ($S_{\text{int}}$) and peak flux values ($S_{\text{peak}}$), respectively. Using the criteria where a source will be considered an extended source if $\ln({S_{\text{int}}}/{S_{\text{peak}}}) > 3{\sigma_\text{R}}$ \citep{2019PASA...36....4F}, we estimated that the resolved sources (steel-blue stars) are 1966, and the compact sources (navy hexagons) are 3775. Figure~\ref{fig:Fig5} presents the source classification of our catalog.

\subsection{Comparison with prior Radio catalogs} \label{sec:comparison}

We compared this uGMRT catalog with the ASKAP's EMU survey of the GAMA-23 field at 887.5\,MHz by \cite{2022MNRAS.512.6104G}. Further, we carried out comparisons with the 1.4\,GHz NVSS survey catalog by \cite{1998AJ....115.1693C}, the 150\,MHz TGSS catalog by \cite{2017A&A...598A..78I}, the 200\,MHz GLEAM-X by \cite{2022PASA...39...35H} and the 300\,MHz GLEAM by \cite{2025PASA...42..158D}.\\ 
We have identified equivalents to the uGMRT sources in other catalogs using the search radius mentioned in Table~\ref{table:tab3}. Each catalog used for comparison has its own observational sensitivity and completeness, which determine its flux density limit ($S_\text{limit}^{\dagger}$). The flux limit at 325 MHz, $S_\text{{cut}}^{\text{325}}$, is used to select sources. We have selected sources with flux densities above $S_\text{limit}^{\dagger}$, scaled to 325\,MHz with $S_\nu \propto \nu^{\alpha}$, where $\alpha=-0.7$. The information of the chosen catalogs for comparison is given in Table~\ref{table:tab3}.
%%%%%%%%%%%%%%%%%%%%%%%%%%%%%%%%%%%%%%%%%%%%%%%%%%%%%%%%%%%%%%%%%%%%%%%%%%%%%%%%%%%%%%%
\begin{table*}
\caption{The selected catalogs for comparison with their frequency, resolution, flux limit before and after scaling at 325\,MHz (assumption $\alpha = -0.7$). Median positional offsets in R.A. and decl. of the uGMRT 325\,MHz catalog sources relative to other catalogs are presented, along with their 16th and 84th percentile errors. The matched sources corresponding to other catalogs are mentioned in the last column.}
\centering
\begin{tabular}{ccccccccc}
\hline
Catalog & Frequency & Resolution & $S_\text{limit}^{\dagger}$ & $S_\text{{cut}}^{\text{325}}$ & Cross-match Radius & $\delta_\text{R.A.,median}$ & $\delta_\text{Decl.,median}$ & Match \\ [0.5ex]
 & (MHz) & (arcsec) & (mJy) & (mJy) & (arcsec) & (arcsec) & (arcsec)  & \\ [1ex]
\hline\hline
uGMRT & 325 & 15.8 & 0.55 & 0.55 & -- & -- & -- & -- \\ [0.5ex]
NVSS & 1400 & 45 & 2.0 &  6.94 & 20 & -0.018$^{+0.44}_{-0.48}$ & -0.072$^{+0.68}_{-0.58}$ & 206\\ [0.5ex]
TGSS & 150 & 25 & 17.5 &  10.2 & 20 & -0.120$^{+0.75}_{-0.61}$ & 0.038$^{+1.01}_{-0.96}$ & 377 \\ [0.5ex]
EMU &   887.5 & 10 & 0.2 &  0.4 & 11 & 0.007$^{+0.47}_{-0.48}$ & 0.246$^{+0.76}_{-0.71}$ & 1583 \\ [0.5ex]
GLEAM-X & 200 & 45 & 3.5 &  2.5 & 70 & -0.005$^{+0.62}_{-0.51}$ & 0.098$^{+0.89}_{-0.88}$ & 644  \\
\hline
\end{tabular}
\label{table:tab3}
\tablecomments{$S_\text{limit}^{\dagger}$ - Flux density limit for above corresponding catalog.}
\end{table*}
%%%%%%%%%%%%%%%%%%%%%%%%%%%%%%%%%%%%%%%%%%%%%%%%%%%%%%%%%%%%%%%%%%%%%%%%%%%%%%%%%%%%
%%%%%%%%%%%%%%%%%%%%%%%%%%%%%%%%%%%%%%%%%%%%%%%%%%%%%%%%%%%%%%%%%%%%%%%%%%%%%%%%%%%%

\subsubsection{Flux Density Offset} \label{sec:fluxoffset} 

\begin{figure}[t]
\includegraphics[width=3.35in]{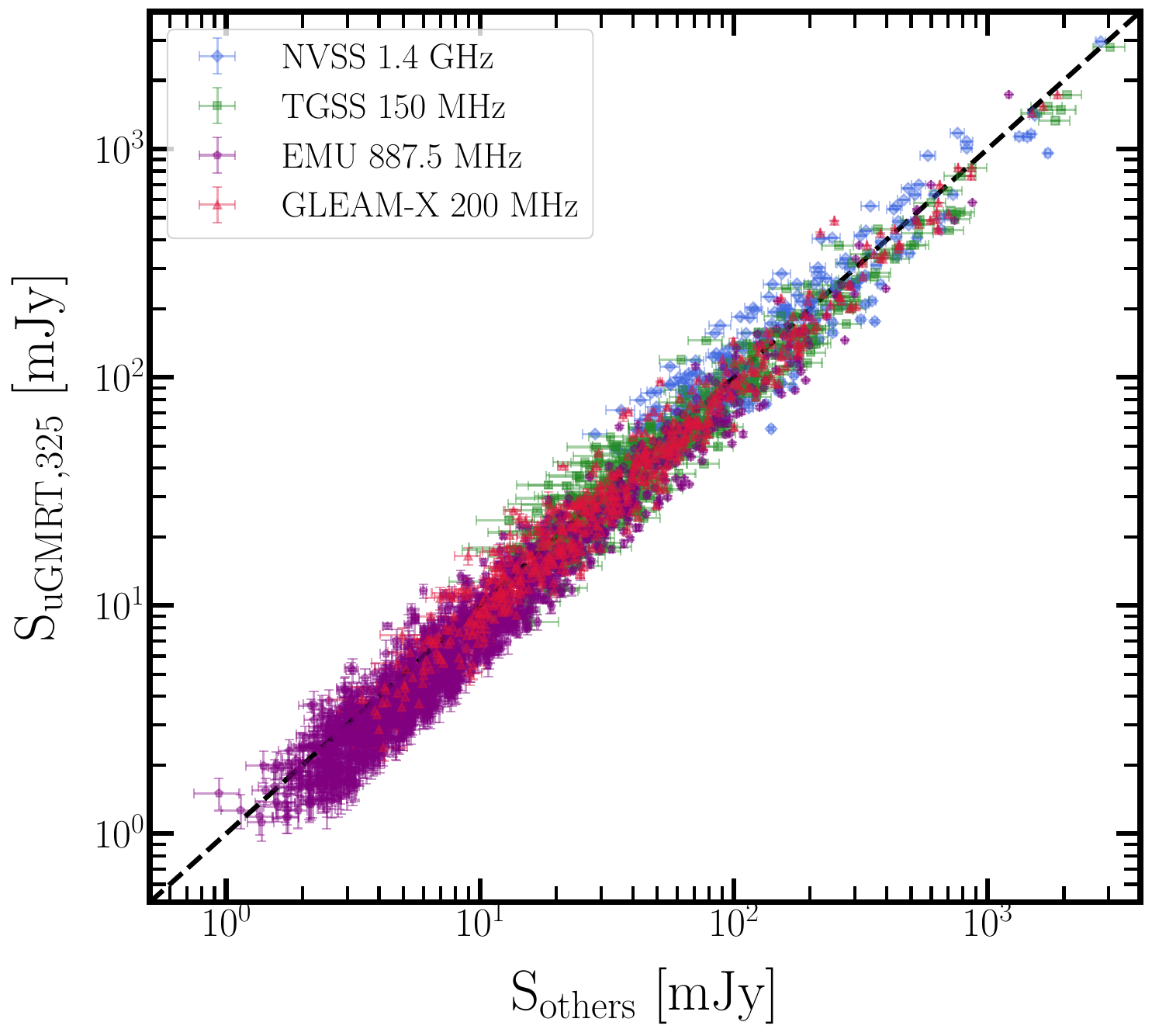}
\caption{Comparison of integrated flux densities of compact, isolated sources from the uGMRT 325\,MHz catalog with those from other catalogs, scaled to 325\,MHz, including 1.4\,GHz NVSS survey (blue), 150\,MHz TGSS (green), 200\,MHz GLEAM-X (red), and 887.5\,MHz EMU (purple). The black dashed line represents $S_{\text{uGMRT}}/S_{\text{others}} = 1$.}
\label{fig:Fig6}
\end{figure}

In this section, the flux density of the source catalog is compared with that of the other catalogs to estimate the flux offset. We used the Scaife-Heald flux scale \citep{2012MNRAS.423L..30S}, similar to the 150\,MHz TGSS catalog \citep{2017A&A...598A..78I}. For other catalogs we have converted them into this flux scale. We used the criteria followed by \cite{2025MNRAS.544.3617S} to select sources for comparison. We have selected the high-SNR sources (10$\sigma$) and compact sources with a size less than the higher resolution in the other catalog. We have also selected isolated sources for which the distance between two sources is greater than 2 times the PSF in the lower-resolution catalog. The selected sources that met this criterion were considered for analysis and are listed as matched sources in Table~\ref{table:tab3}. The cross-matching radius was determined based on the angular resolution of the comparison catalogs. The adaptive radius is $0.5-1\,{\times}$\,PSF to balance completeness and minimize spurious associations, while $1.5\,{\times}$\,PSF for GLEAM-X. Flux density ratios of uGMRT 325\,MHz catalog with other catalogs ($S_{325 \text{MHz}}/S_{\text{others}}$) after scaling were estimated with ${\alpha}\,{\sim} -0.7$ (assumption). In this ratio, ``others'' refers to the catalogs used for comparison. The derived medians of the flux ratios are found to be:  0.91$^{+0.22}_{-0.36}$ (NVSS 1.4\,GHz), 0.80$^{+0.24}_{-0.18}$ (ASKAP EMU 887.5\,MHz), 0.88$^{+0.12}_{-0.15}$ (GLEAM 300\,MHz), 0.90$^{+0.13}_{-0.20}$ (GLEAM-X 200\,MHz), and 0.91$^{+0.18}_{-0.26}$ (TGSS 150\,MHz) with 16th and 84th percentile errors. Figure~\ref{fig:Fig6} presents the flux offset values of selected sources compared to other catalogs. The median values are near 1, as shown by the black dashed line.
%%%%%%%%%%%%%%%%%%%%%%%%%%%%%%%%%%%%%%%%%%%%%%%%%%%%%%%%%%%%%%%%%%%%%%%%%%%%%%%%%%
%%%%%%%%%%%%%%%%%%%%%%%%%%%%%%%%%%%%%%%%%%%%%%%%%%%%%%%%%%%%%%%%%%%%%%%%%%%%%%%%%%%%

\subsubsection{Positional Offset} \label{sec:pos-off} 

The positional accuracy of uGMRT source catalog has been estimated by comparing our catalog with the NVSS 1.4\,GHz, TGSS 150\,MHz, GLEAM-X 200\,MHz, ASKAP EMU survey catalogs at 887.5\,MHz. We followed the same criteria as in Section~\ref{sec:fluxoffset} to select sources. The 1.4\,GHz NVSS catalog \citep{1998AJ....115.1693C} has the highest frequency among selected catalogs, resulting in less ionospheric fluctuation and thus provides better positional accuracy ($<1''$). We used the method mentioned in \cite{2016MNRAS.460.2385W}:
%%%%%%%%%%%%%%%%%%%%%%%%%%%%%%%%%%%%%%%%%%%%%%%%%%%%%%%%%%%%%%%%%%%%%%%%%
\begin{equation}
\begin{aligned}
\label{eqn:4}
\delta_\mathrm{RA} & = \mathrm{RA_{uGMRT} - RA_{NVSS}} \\
\delta_\mathrm{DEC} & = \mathrm{DEC_{uGMRT} - DEC_{NVSS}}
\end{aligned}
\end{equation}
%%%%%%%%%%%%%%%%%%%%%%%%%%%%%%%%%%%%%%%%%%%%%%%%%%%%%%%%%%%%%%%%%%%%%%%%%
%%%%%%%%%%%%%%%%%%%%%%%%%%%%%%%%%%%%%%%%%%%%%%%%%%%%%%%%%%%%%%%%%%%%%%%%%%%%%%%%%%%%
\begin{figure}
\includegraphics[width=3.35in]{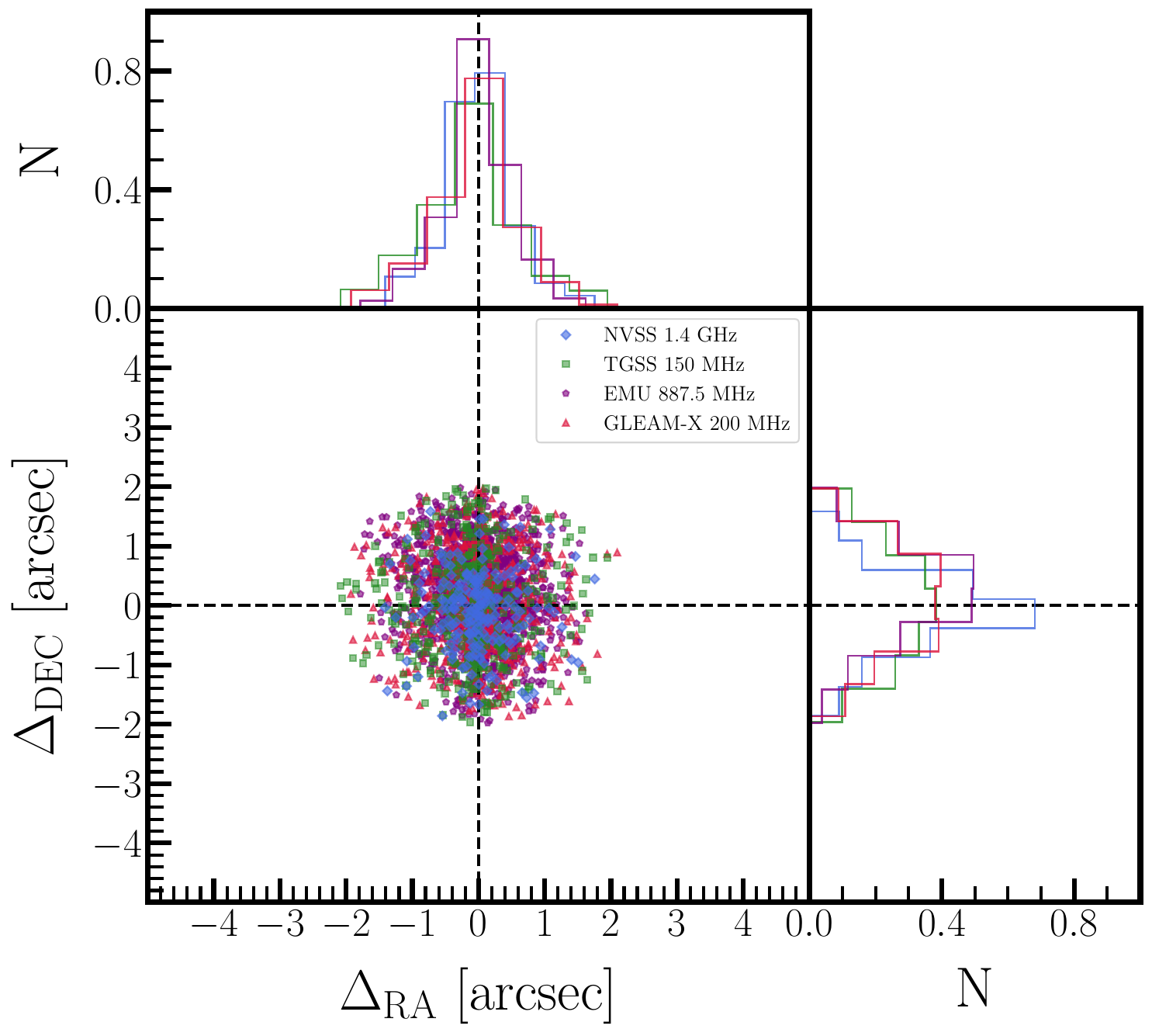}
\caption{Positional offset (R.A. $\&$ Decl.) in the sources of the uGMRT catalog at 325\,MHz in comparison with the 1.4\,GHz NVSS (blue), 150\,MHz TGSS (green), 200\,MHz GLEAM-X (red) and 887.5\,MHz ASKAP-EMU (purple) observational surveys.}
\label{fig:Fig7}
\end{figure}
%%%%%%%%%%%%%%%%%%%%%%%%%%%%%%%%%%%%%%%%%%%%%%%%%%%%%%%%%%%%%%%%%%%%%%%%%%%%%%%%%%%%%%%%%%
Table~\ref{table:tab3} shows the values of median offset derived from different catalogs used. The median deviations in R.A. and decl., evaluated using the NVSS catalog, are $0.018''$ and $−0.072''$, respectively. Figure~\ref{fig:Fig7} presents histograms of positional offset in RA and declination for this catalog in comparison with other selected catalogs. The offset measured from the ASKAP EMU GAMA-23 field %and ASKAP RACS survey 
catalog (see Table~\ref{table:tab3}) is minimal in comparison to the image cell size of $1.9''$ of uGMRT. We have not applied corrections to the source positions in our uGMRT catalog, as the resolution of our catalog ($15.8''$) is better than the NVSS catalog ($45''$).

\subsubsection{Spectral Index Distribution}
%%%%%%%%%%%%%%%%%%%%%%%%%%%%%%%%%%%%%%%%%%%%%%%%%%%%%%%%%%%%%%%%%%%%%%%%%%%%%%%%%%%%%%%%%%%%%%%%%%%%
\begin{figure}
    \includegraphics[width=3.35in]{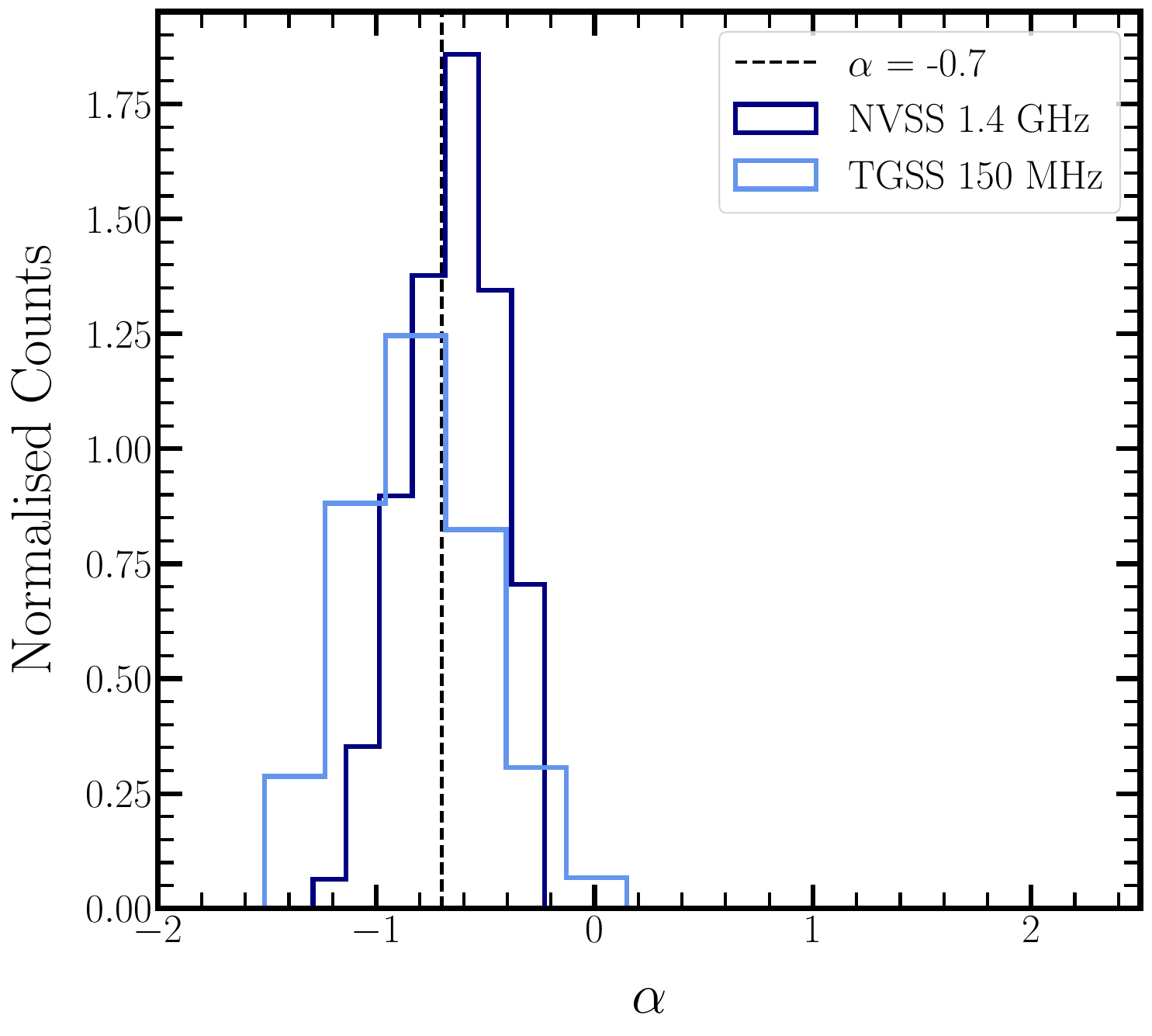}
     \caption{The normalized distribution of spectral indices derived from cross-matching with the 1.4\,GHz VLA-NVSS (navy) and 150\,MHz uGMRT-TGSS (royal blue) surveys is shown as solid lines, using a matching radius of $20''$. The assumed power-law spectral index of $\alpha = -0.7$ is indicated by a dashed line in black.}
     \label{fig:Fig8}
 \end{figure}
%%%%%%%%%%%%%%%%%%%%%%%%%%%%%%%%%%%%%%%%%%%%%%%%%%%%%%%%%%%%%%%%%%%%%%%%%%%%%%%%%%%%
To characterize the spectral properties in the GAMA-23 deep field, we have calculated the flux densities of sources from this uGMRT catalog and compared them with other available catalogs. We have derived two point-spectral indicies using NVSS 1.4\,GHz \citep{1998AJ....115.1693C} and TGSS 150\,MHz \citep{2017A&A...598A..78I} catalog. We have used the source selection norm as follows in Section~\ref{sec:fluxoffset} and ~\ref{sec:pos-off}. We premised the synchrotron power-law distribution in the form of $S_\nu \propto \nu^{\alpha}$. The spectral index, $\alpha$, is calculated for 206 and 377 matched sources in this catalog using the NVSS and TGSS catalogs. Figure~\ref{fig:Fig8} presents the spectral indices $\alpha$ of matched sources in terms of histograms. The estimated median spectral values, ${\alpha}_\text{median}$ are ${-0.65}^{+0.23}_{-0.19}$ (NVSS 1.4\,GHz) and ${-0.81}^{+0.28}_{-0.33}$ (TGSS 150\,MHz) with errors for the 16th and 84th percentiles. A comparison of our catalog with the NVSS-TGSS spectral index catalog \citep{2018MNRAS.474.5008D} shows comparable spectral index distribution. For 341 matched sources, the median spectral index is $-0.77\,{\pm}\,0.08$ for uGMRT catalog and $-0.64\,{\pm}\,0.03$ for the NVSS-TGSS catalog. These are consistent with the median spectral index obtained from individual NVSS and TGSS catalog comparisons. It is also consistent with synchrotron dominated sources, for which ${\alpha}\,{\sim} −0.5$ to $−0.7$ \citep{1992ARA&A..30..575C}, with steeper spectra tracing older electron populations and diffuse emission. Similar median spectral indices (${\alpha}\,{\sim}−0.6$ to $−0.7$) have been reported in deep radio surveys \citep{2009MNRAS.397..281I, 2012MNRAS.426.2342H, 2019MNRAS.490..243C, 2025MNRAS.544.3617S}.

\section{Source Count} \label{sec:count}

%%%%%%%%%%%%%%%%%%%%%%%%%%%%%%%%%%%%%%%%%%%%%%%%%%%%%%%%%%%%%%%%%%%%%%%%%%%%%%%%%%%
\begin{figure}
\includegraphics[width=3.35in]{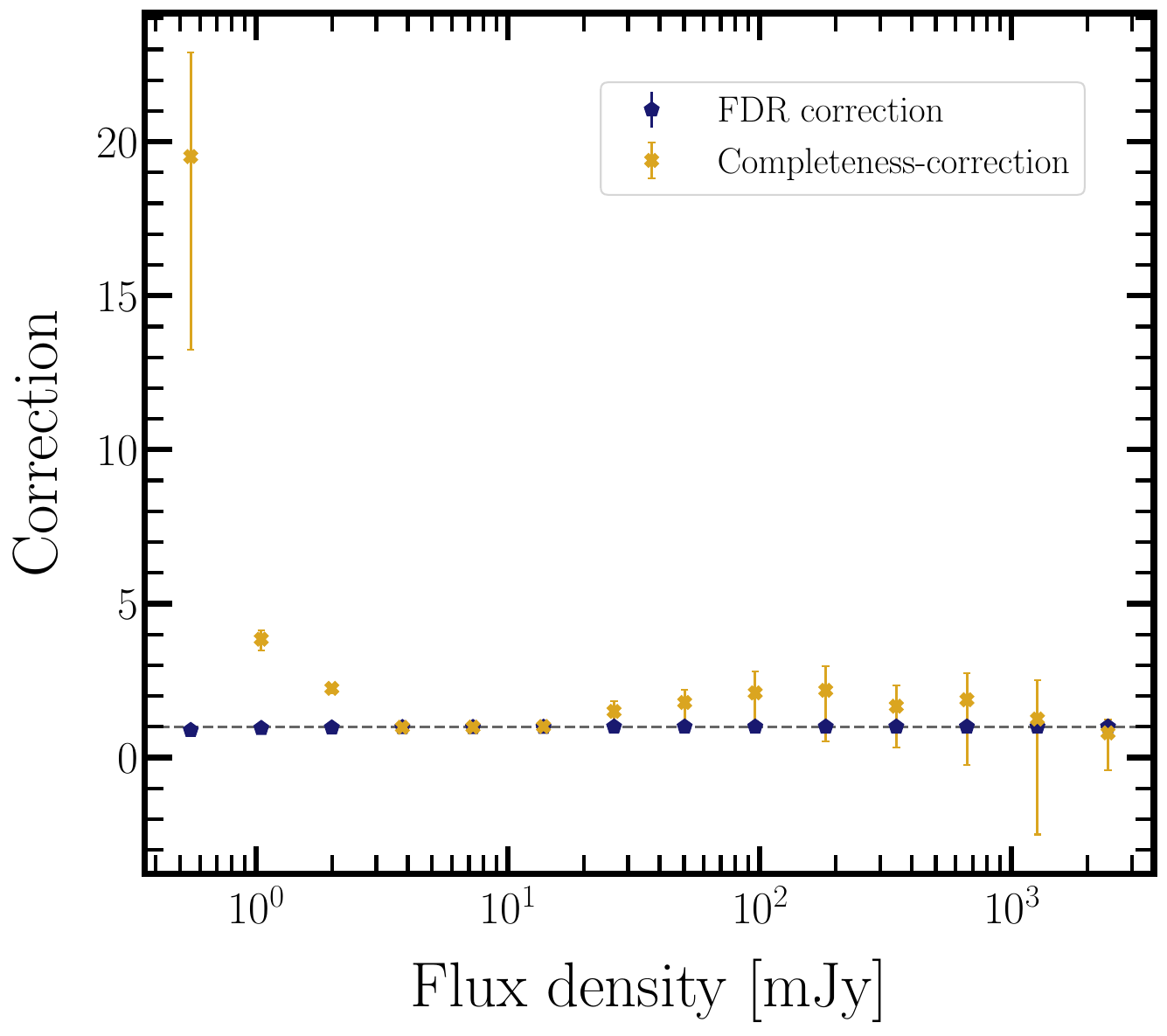}
\caption{Correction factors in each flux bin for FDR (shown in blue) and completeness (shown in golden) with integrated flux density.}
\label{fig:Fig9}
\end{figure}
%%%%%%%%%%%%%%%%%%%%%%%%%%%%%%%%%%%%%%%%%%%%%%%%%%%%%%%%%%%%%%%%%%%%%%%%%%%%%%%%%%%%
Source counts are defined as the number of sources within a specified flux density bin per unit solid angle on the sky. The source counts are multiplied by S$^{2.5}$ to derive the Euclidean normalized differential source counts \citep{2013MNRAS.432.2625H}. At low frequencies, source counts are used to distinguish the behavior of SFGs and AGNs, and to study the fundamental mechanisms driving the behavior of radio sources \citep{2001A&A...365..392P}. The flux density, in general, is modeled as $dN/dS\,{\propto}\,S^{-1.6}$, a single power-law distribution.  \\
The GAMA-23 field consists of enriched multiwavelength data and low Galactic synchrotron contamination, making it an ideal field for studying faint sources. We have estimated the normalized differential source counts for the GAMA-23 field, using uGMRT flux densities derived from the PyBDSF catalog. We measured the differential source counts down to sub-mJy (i.e., 0.5\,mJy) at 325\,MHz. However, output from PyBDSF can contain biases due to false-detection rate (FDR) and catalog incompleteness. The presence of these biases can significantly impact the precision of the true value of source counts, especially at faint flux density bins. We computed two corrections, FDR and incompleteness, for the normalized source counts. These corrections account for the effects of visibility area and resolution on-source detection. The corrections calculation and application to the normalized source count distribution are detailed below.

\subsection{False-detection Rate}

We accounted for the FDR in the generated catalog by considering the source detection over the negative image (i.e., −1 $\times$ image). This method considers the contamination of the noise in the final catalog. This method depends on the assumption that the noise in the image is symmetric.
Therefore, we generated an inverted catalog from this image using the same PyBDSF parameters as applied to the original image (see Section~\ref{sec:catalog}). A total of 126 negative peaks below ${-5}{\sigma}$ were detected, which represent false detections arising from noise. These negative sources were binned into 14 logarithmic flux density bins. The number of negative detections in each bin was then compared with the corresponding counts from the original image. 
%%%%%%%%%%%%%
\begin{figure*} 
\centering
 \makebox[\textwidth]{\includegraphics[width=0.8\paperwidth]{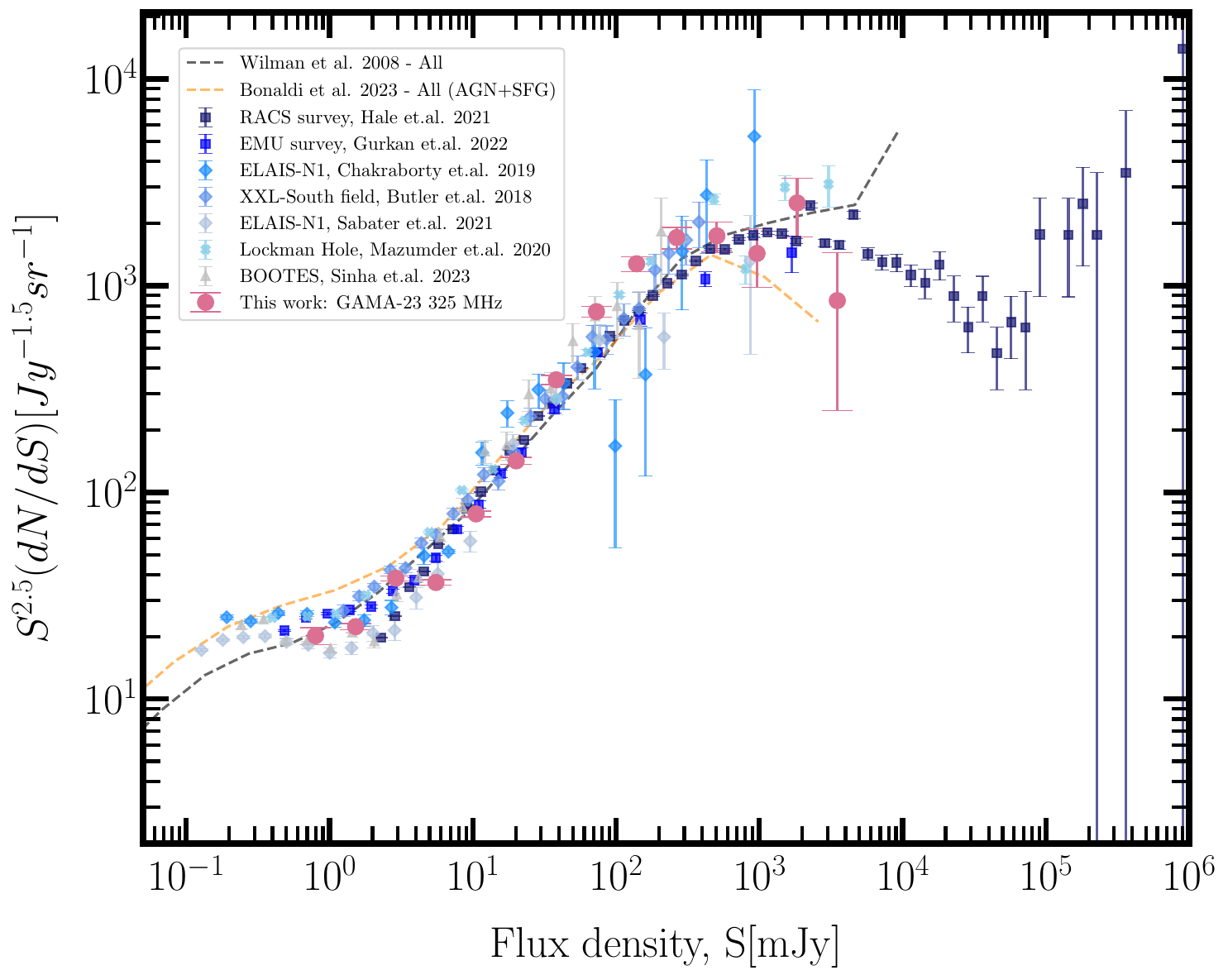}}
\caption{The Euclidean-normalized differential source counts at 325\,MHz for the GAMA-23 field (pink circles) are shown after applying FDR and incompleteness corrections. Model predictions from the S$^3$-SKADS \citep{2008MNRAS.388.1335W} and T-RECS II \citep{2023MNRAS.524..993B} simulations are indicated as black and orange dashed lines, respectively. The results are also compared with other observational datasets covering the same sky region, such as ASKAP RACS survey at 887.5\,MHz \citep{2021PASA...38...58H} (navy squares), EMU (GAMA-23) survey 887.5\,MHz \citep{2022MNRAS.512.6104G} (blue squares). Comparisons with other fields include ELAIS-N1 - 400\,MHz uGMRT \citep{2019MNRAS.490..243C} (dodger blue diamonds), LoTSS 150\,MHz LOFAR \citep{2021A&A...648A...5M} (light steel blue diamonds), XXL$-$South field 2.1\,GHz ATCA \citep{2018A&A...620A...3B} (cornflower-blue diamonds), Lockman Hole 325\,MHz GMRT \citep{2020MNRAS.495.4071M} (skyblue crosses), and Bo\"{o}tes 400\,MHz uGMRT \citep{2023MNRAS.525.5311S} (silver triangles).
}
\label{fig:Fig10}
\end{figure*}
%%%%%%%%%%%%%%%%%%%%%%%%%%%%%%%%%%%%%%%%%%%%%%%%%%%%%%%%%%%%%%%%%%%%%%%%%%%%%%%%%%%%
The number of real sources in the k$^\text{th}$ flux density bin of source counts presented in \cite{2019A&A...622A...4H} by the equation:
%%%%%%%%%%%%%%%%%%%%%%%%%%%%%%%%%%%%%%%%%%%%%%%%%%%%%%%%%%%%%%%%%%%%%%%%%%%%%%%%%%%%
\begin{equation}
\begin{aligned}
\label{eqn:6}
f_{\text{real}, \text{k}} = \frac{N_{\text{catalog}, \text{k}} - N_{\text{inv}, \text{k}}}{N_{\text{catalog}, \text{k}}}
\end{aligned}
\end{equation}
%%%%%%%%%%%%%%%%%%%%%%%%%%%%%%%%%%%%%%%%%%%%%%%%%%%%%%%%%%%%%%%%%%%%%%%%%%%%%%%%%
where N$_\text{inv,k}$ and N$_\text{catalog,k}$ denote the number of detected sources in the k$^\text{th}$ flux density bin of the negative and original images, respectively. We used Poissonian statistics to estimate the uncertainties. The estimated fraction was scaled by the number of sources in each flux density bin to obtain completeness correction factors for the original catalog. Figure~\ref{fig:Fig9} presents the FDR correction in blue pentagons for each bin.

\subsection{Completeness}\label{sec:completenss}

The incompleteness in the catalog arises due to non-uniform noise in the image. Though when there are different pointings and a mosaic has been created, the overlapping of primary beams promises to compensate for the attenuation of one pointing by adjacent pointings. This results in almost uniform noise distribution and consistent source detection across the field. However, the source finding algorithm has limitations. Also, biases such as the Eddington bias \citep{1913MNRAS..73..359E}, and resolution bias \citep{2006A&A...457..517P} are primary reasons to introduce incompleteness. Eddington bias leads to overestimation of sources near the detection limit when noise scatters more faint sources above the detection threshold than bright sources below it. Resolution bias occurs when extended sources have their flux spread over a large area, resulting in low peak flux densities that can fall below the detection threshold. Therefore, causing incompleteness in the source catalog.

\begin{deluxetable*}{ccccccccc}
\tablecaption{The estimated Euclidean normalized differential source counts of the GAMA-23 field at 325\,MHz. The columns below are associated with the bins of flux density, the central estimate of the flux density bin, normalized raw source counts, correction factors for FDR and completeness, and normalized corrected source counts.}
\label{table:tab4}
\tablehead{
\colhead{S$_{\text{range}}$} & \colhead{S$_{\text{mid}}$} & \colhead{N} & \colhead{S$^{2.5}$(dN/dS)} & \colhead{FDR} & \colhead{Completeness} & \colhead{Corrected S$^{2.5}$(dN/dS)} \\ 
\colhead{(mJy)} & \colhead{(mJy)} & \colhead{} & \colhead{(Jy$^{1.5}$sr$^{-1}$)} & \colhead{} & \colhead{} & \colhead{(Jy$^{1.5}$sr$^{-1}$)} 
}
\startdata
0.550\,--\,1.048 & 0.799 & 116 & 1.166 $\pm$ 0.108 & 0.89 $\pm$ 0.11 &  19.52$^{+3.39}_{-6.28}$  & 20.261 $\pm$ 1.881 \\
1.048\,--\,1.999 & 1.524 & 764 & 6.130 $\pm$ 0.222 & 0.95 $\pm$ 0.05 & 3.85$^{+0.30}_{-0.37}$ & 22.396 $\pm$ 0.810 \\
1.999\,--\,3.810 & 2.904 & 1260 & 17.645 $\pm$ 0.497 & 0.97 $\pm$ 0.03 & 2.25$^{+0.14}_{-0.16}$ & 38.443 $\pm$ 1.083 \\
3.810\,--\,7.262 & 5.536 & 1127 & 38.095 $\pm$ 1.135 & 0.98 $\pm$ 0.02 & 0.98$^{+0.10}_{-0.13}$ & 36.661 $\pm$ 1.092 \\
7.262\,--\,13.843 & 10.552 & 927 & 81.183 $\pm$ 2.666 & 0.98 $\pm$ 0.02 & 0.99$^{+0.17}_{-0.11}$ & 78.684 $\pm$ 2.584 \\
13.843\,--\,26.387 & 20.115  & 620 & 142.572 $\pm$ 5.726 & 0.99 $\pm$ 0.01 & 1.01$^{+0.12}_{-0.17}$ & 142.134 $\pm$ 5.708 \\
26.387\,--\,50.299 & 38.343  & 386 & 233.600 $\pm$ 11.890 & 1.00 $\pm$ 0.00 & 1.50$^{+0.32}_{-0.38}$ & 350.400 $\pm$ 17.835 \\
50.299\,--\,95.881 & 73.090 & 263 & 418.887 $\pm$ 25.830 & 1.00 $\pm$ 0.00 & 1.79$^{+0.41}_{-0.65}$ & 748.969 $\pm$ 46.183 \\
95.881\,--\,182.769 & 139.325 & 145 & 607.806 $\pm$ 50.476 & 1.00 $\pm$ 0.00  & 2.10$^{+0.70}_{-0.93}$ & 1276.393 $\pm$ 105.999 \\
182.769\,--\,348.396 & 265.583 & 71 & 783.271 $\pm$ 92.957 & 1.00 $\pm$ 0.00  & 2.18$^{+0.79}_{-1.65}$ & 1709.880 $\pm$ 202.925 \\
348.396\,--\,664.117 & 506.257 & 36 & 1045.232 $\pm$ 174.205 & 1.00 $\pm$ 0.00  & 1.67$^{+0.67}_{-1.33}$ & 1742.402 $\pm$ 290.400 \\
664.117\,--\,1265.947 & 965.032 & 10 & 764.129 $\pm$ 241.639 & 1.00 $\pm$ 0.00  & 1.88$^{+0.88}_{-2.13}$ & 1432.742 $\pm$ 453.073 \\
1265.947\,--\,2413.164 & 1839.556 & 10 & 2011.053 $\pm$ 635.951 & 1.00 $\pm$ 0.00  & 1.25$^{+1.25}_{-3.75}$ & 2513.816 $\pm$ 794.938 \\
2413.164\,--\,4600.000 & 3506.582 & 2 & 1058.547 $\pm$ 748.506 & 1.00 $\pm$ 0.00  & 0.80$^{+0.43}_{-1.20}$ & 846.837 $\pm$ 598.804 \\
\enddata
\end{deluxetable*}
%%%%%%%%%%%%%%%%%%%%%%%%%%%%%%%%%%%%%%%%%%%%%%%%%%%%%%%%%%%%%%%%%%%%%%%%%%%%%%%%%%%%

We accounted for incompleteness in source counts due to biases and conducted simulations using {\tt AERES} from the open-source algorithm {\tt AEGEAN}\footnote{\url{https://github.com/PaulHancock/Aegean}} \citep{2012ascl.soft12009H, 2018PASA...35...11H}. We randomly injected 1500 sources into the RMS map generated using PyBDSF \citep[similar as][]{2000A&AS..146...41P, 2025MNRAS.544.3617S}. The flux density of sources follows a differential slope close to $dN/dS \propto S^{-1.6}$ \citep{1984ApJ...287..461C, 2008MNRAS.388.1335W, 2013A&A...549A..55W}. The range of flux densities of injected sources was $0.5- 4600$\,mJy. There were 600 extended sources of sizes $4.6''-8.3''$ and the remaining were point-like sources with $\theta_{\text{major}} \geq 15.8''$, similar to source classification in the original catalog. We have performed 100 simulation realizations to consider the impact on visibilityarea and constraints on-source confusion \citep{2019PASA...36....4F, 2019A&A...622A...4H}. We have extracted catalogs from 100 images using PyBDSF with the same parameters as mentioned in Section~\ref{sec:catalog}. The sources in these extracted catalogs were aligned with the original catalog binning. For completeness, the correction factor in the k$^\text{th}$ flux bin is as follows:
%%%%%%%%%%%%%%%%%%%%%%%%%%%%%%%%%%%%%%%%%%%%%%%%%%%%%%%%%%%%%%%%%%%%%%%%%%%%%%%%%%%%%%%%%%%%%%%%%%%%
\begin{equation}
\begin{aligned}
\label{eqn:7}
\text{Correction}_\text{k} = \frac{N_{\text{injected, k}}}{N_{\text{recovered, k}}}
\end{aligned}
\end{equation}
%%%%%%%%%%%%%%%%%%%%%%%%%%%%%%%%%%%%%%%%%%%%%%%%%%%%%%%%%%%%%%%%%%%%%%%%%%%%%%%%%%%%
where N$_\text{injected,k}$ and N$_\text{recovered,k}$ indicate the number of injected and restored sources within the kth bin after eliminating pre-simulation sources \citep{2019A&A...622A...4H}. Figure~\ref{fig:Fig9} represents the FDR and completeness correction factor for each flux density bin. The correction factor in each flux density bin was derived from the median value across 100 realizations, along with the corresponding errors at the 16th and 84th percentiles.

\subsection{Differential Source Count} \label{sec:dN/dS}

Using our extracted catalog and corrections for FDR and completeness with 100 realizations, we have estimated Euclidean-normalized source counts in units of Jy$^{-1.5}$sr$^{-1}$. We multiplied the correction factors obtained from FDR and completeness by the normalized raw source counts of each bin. The noise across the image varies (lower panel of Figure~\ref{fig:Fig2}). Thus, the effective area over which a source can be detected depends on its flux density \citep{1985ApJ...289..494W}. Bright sources can be identified across the image, while faint sources are detectable only in low-noise regions. Therefore, an effective area correction was applied separately to each flux density bin. The flux densities of sources were binned into 14 logarithmic bins in the range $0.5-4600$\,mJy and associated errors were estimated using Poisson statistics. The normalized differential source counts (raw and corrected) for each flux density bin, along with correction factors, are presented in Table~\ref{table:tab4}. We present the normalized source count in Figure~\ref{fig:Fig10} after implementing the corrections from FDR and incompleteness. We considered the total flux densities of both compact and extended sources to derive the source count.
\begin{figure*}
  \centering
    \includegraphics[width=\linewidth]{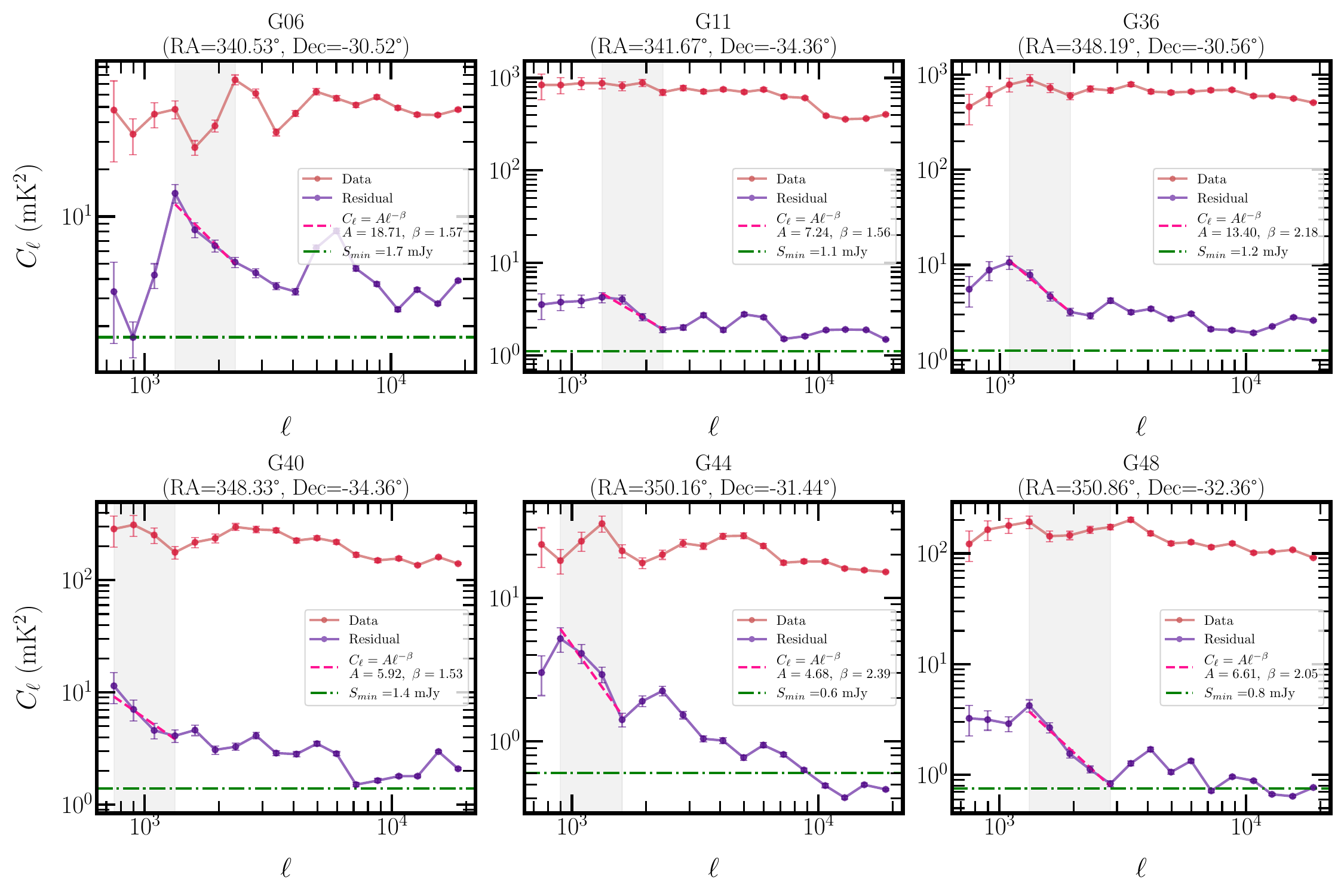}
  \caption{Angular power spectrum ($C_{\ell}$) with $1{\sigma}$ error bars (in red and purple) plotted as a function of angular multipole ($\ell$) for the visibility-based estimator \citep[TGE;][]{2014MNRAS.445.4351C} at frequency 325\,MHz for six target pointings at different latitudes of the GAMA-23 field. The vertical gray shadow indicates the range of ${\ell}$ values used to fit the power law, and the best-fitting model $C_{\ell} = A(1000/{\ell})^{{\beta}}$ is shown in magenta. The dashed dotted line (in green) represents the fitting prediction \citep[i.e., ${S_\text{min}=5{\sigma}}$;][]{2008MNRAS.385.2166A}.  The values of magnitude A, power-law index ${\beta}$ and ${\ell}$ range for APS are given in Table \ref{table:tab5} for Band-3.} 
  \label{fig:Fig11}
\end{figure*}
%%%%%%%%%%%%%%%%%%%%%%%%%%%%%%%%%%%%%%%%%%%%%%%%%%%%%%%%%%%%%%%%%%%%%%%%%%
For cross verification, we compared these with other previously observed source counts and simulations. We used ASKAP EMU GAMA-23 field at 887.5\,MHz \citep{2022MNRAS.512.6104G} and scaled to 325\,MHz using a spectral index, ${\alpha} = {-0.7}$. We have also compared our catalog with other uGMRT catalogs of different fields and at different frequency bands, such as the ELAIS-N1 field at 400\,MHz \citep{2019MNRAS.490..243C} and LoTSS at 150\,MHz \citep{2021A&A...648A...5M}, XXL$-$South field at 2.1\,GHz \citep{2018A&A...620A...3B}, 325\,MHz Lockman Hole field \citep{2020MNRAS.495.4071M} and 400\,MHz Bo\"{o}tes field \citep{2023MNRAS.525.5311S}. The estimated corrected source counts are presented with pink circles in Figure~\ref{fig:Fig10}. These source counts are consistent with the ASKAP EMU, RACS and other surveys with different fields yet same frequency band from uGMRT \citep{2019MNRAS.490..243C, 2020MNRAS.495.4071M, 2023MNRAS.525.5311S}. This work is close to the lowest bin of the ASKAP EMU survey, i.e., 0.2\,mJy. We have also considered simulated catalogs for comparison, such as $S^3$-SKADS \citep{2008MNRAS.388.1335W} and T-RECS II \citep{2023MNRAS.524..993B}. The sources in this catalog follow the trend well with both the simulations and ASKAP (EMU and RACS) catalogs. 

\section{Power Spectrum of Diffuse Emission}\label{sec:DGSE_APS}

One of the main obstacles to detecting the 21 cm signal is the residual from point source subtraction and foregrounds \citep{2005ApJ...619..678M, 2010ApJ...724..526D}. Extensive work on simulation in foreground modeling has been done \citep{2008MNRAS.385.2166A, 2008MNRAS.389.1319J, 2010MNRAS.409.1647J,  10.1093/mnras/sty2348}. However, the foregrounds behave differently at different frequencies across patches of sky. Thus, their study in those fields is necessary to model their true nature \citep{2017MNRAS.470L..11C, 2019MNRAS.487.4102C, 2020MNRAS.495.4071M, 2022MNRAS.510.2011B, 2022A&A...662A..97G, 2025PASA...42..103C, 2025MNRAS.544.3617S}. The foregrounds are $4-5$ orders of magnitude dominant and smooth, unlike the 21 cm brightness fluctuation. Thus, we can fit DGSE in the form of a power law for spectral and spatial characteristics. \\
\cite{2019MNRAS.490..243C} presented the break in DGSE single power law at 405\,MHz in the ELAIS-N1 deep field. In this work, we present the spatial characterization of the DGSE in the southern sky. The calibrated data used in this analysis is generated in the calibration procedure discussed in Section~\ref{sec:calibration}. We present DGSE in terms of power law ($C_{\ell} = A(1000/{\ell})^{{\beta}}$) using a visibility-based estimator, namely the tapered gridded estimator \citep[TGE;][]{2014MNRAS.445.4351C, 2016MNRAS.459..151C, 2017MNRAS.470L..11C}. We used calibrated visibility data, both with and without point sources, to estimate the APS in the target pointings. The catalog analysis in Section~\ref{sec:catalog} validates the flux density and astrometric accuracy of the point sources in the calibrated data. In addition, the CLEAN model generated with WsClean was used to produce the residual visibilities for DGSE analysis. We have used $100-3000$\,m baseline\footnote{\url{https://www.gmrt.ncra.tifr.res.in/doc/GMRT_specs.pdf}} range (corresponds to $1'-32'$ angular scale) to estimate the APS for each pointing. We did not apply tapering \citep[${\theta}_{w} = f{\theta}_0$, $f$; see][]{2014MNRAS.445.4351C} as low multipoles ${\ell}$ are sensitive to large angular scale. They are, in general, affected by primary beam uncertainty \citep{2017MNRAS.470L..11C, 2020MNRAS.494.1936C}. \\
We have selected six different target pointings at different latitudes to understand the nature of DGSE over the GAMA-23 field. The multipole range ($\ell_\mathrm{min}-\ell_\mathrm{max}$) has been selected where the residual APS is described by single power law for each pointing. The goodness of fit was estimated using the reduced $\chi^2$ value. Figure~\ref {fig:Fig11} shows the APS in the red curve from the point sources containing visibility data, and the residual APS is shown with the purple curve. Both of the curves show $1-{\sigma}$ error bars on the data points. The fitted power law represented with the magenta line. The fitted range of $\ell_{\rm min}$$-$$\ell_{\rm max}$ has been presented within the gray shaded region in each target pointings. The minimum flux density S$_\text{min}$ (in green) presents the 5$\sigma$ value for each target pointings, respectively. The R.A. and Decl. (in degrees) of the target pointings are mentioned above each subplot. The parameter values are shown in the legend. All the parameters A, ${\beta}$, ${\ell}$ range of fitted power law, Galactic latitude $\&$ longitude (in degrees) and reduced ${\chi}^2$ have been mentioned in Table~\ref{table:tab5} in detail.

\begin{deluxetable*}{lcccccc}
\tablecaption{Angular power spectrum estimated using visibility-based estimator \citep{2014MNRAS.445.4351C}. The parameters of the fitted power-law over multipole range $\ell_{\rm min}$$-$$\ell_{\rm max}$ for target pointings with different latitudes across the GAMA-23 field at $325$\,MHz.
\label{table:tab5}}
\tablehead{\colhead{Field ID} & \colhead{R.A. $\&$ Decl.} & \colhead{(${\ell}$, {$|b|$})} & \colhead{$A$} & \colhead{$\beta$} & \colhead{$\ell_{\rm min}$$-$$\ell_{\rm max}$} & \colhead{$\chi^2_{\text{reduced}}$}\\ [-1ex]
 & \colhead{(deg, deg)} & \colhead{(deg, deg)} & (mK$^{2}$) &  &  & }
\startdata
G-06 & ($340.53^{\circ}$, $-30.52^{\circ}$) & ($18.33^{\circ}$, $61.57^{\circ}$) & $18.71 \pm 3.23$ & $1.57 \pm 0.26$ & $1328$$-$$2329$ & 1.09 \\
G-11 & ($341.67^{\circ}$, $-34.36^{\circ}$) & ($10.07^{\circ}$, $62.36^{\circ}$) & $7.24 \pm 1.11$ & $1.56 \pm 0.22$ & $1324$$-$$2332$ & 1.17 \\
G-36 & ($348.19^{\circ}$, $-30.56^{\circ}$) & ($18.03^{\circ}$, $68.16^{\circ}$) & $13.4 \pm 1.94$ & $2.18 \pm 0.28$ & $1091$$-$$1930$ & 0.22\\
G-40 & ($348.33^{\circ}$, $-34.36^{\circ}$) & ($7.93^{\circ}$, $67.78^{\circ}$) & $5.92 \pm 0.62$ & $1.53 \pm 0.53$ & $750$$-$$1326$ & 0.66\\
G-44 & ($350.16^{\circ}$, $-31.44^{\circ}$) & ($15.19^{\circ}$, $69.77^{\circ}$) & $4.68 \pm 0.51$ & $2.39 \pm 0.31$ & $897$$-$$1597$ & 1.80\\
G-48 & ($350.86^{\circ}$, $-32.36^{\circ}$) & ($12.27^{\circ}$, $70.24^{\circ}$) & $6.61 \pm 0.87$ & $2.05 \pm 0.16$ & $1323$$-$$2820$ & 1.22\\
\enddata
\tablecomments{The field IDs correspond to the number IDs mentioned in the Figure~\ref{fig:Fig1}.}
\end{deluxetable*}
In Figure~\ref{fig:Fig11}, the point source containing APS shows Poisson behavior with power of  $20-1000$\,mK$^2$. After point source removal, the power drops by $1-2$ orders and APS begins to be dominated by the DGSE. We fitted the power law curve to the residual APS, which is dominated by the DGSE. The fitted ${\ell}$ range is limited because the unsubtracted point source starts to dominate at large ${\ell}$. The amplitude of DGSE for the target pointings lies within the range $1-20\,{\text{mK}}^2$ with power law fitted over the multipole range $750\,{\leq}\,{\ell}\,{\leq}\,2820$. This result shows that the foreground contamination is minimal in the GAMA-23 field as it lies away from the Galactic plane with Galactic latitude $|b|\,{\sim}\,65^{\circ}$. In this work, the power law index ${\beta} = 1.5-2.4$ is also not much steeper, indicating a significant contribution from small-scale structures \citep{2022MNRAS.509.4923I}. This follows the literature trend of power-law index, i.e., $1.5-3.0$ at low frequencies \citep{2008MNRAS.385.2166A, 2013A&A...558A..72I, 2017MNRAS.470L..11C, 2019MNRAS.490..243C, 2020MNRAS.495.4071M, 2025MNRAS.544.3617S}. Thus, the results from this work (see Table ~\ref{table:tab5}) show that the GAMA-23 deep field is promising to study the \HI\, 21 cm power spectrum due to minimal contamination of foregrounds on large scales.

\section{Conclusions} \label{sec:conclusion}

We present our results from uGMRT observations of the GAMA-23 field at 325\,MHz. We have described the detailed data analysis steps of calibration, mosaicking, catalog extraction, comparison, and source counts. We also presented the first results of foreground characterization in this Galactic latitude ($|b|\,{\sim}\,65^{\circ}$). We performed DD-calibration on 33 hr of uGMRT data, including 50 target pointings, and generated a mosaic image combining all target pointings. The central off-source RMS noise of the mosaic image is 109\,$\mu\text{Jy beam}{^{-1}}$ with a resolution of $15.8''$. The mosaic image has FoV of $14.9^{\circ}\,{\times}\,5.8^{\circ}$ and covers an area of $65.31\,\text{deg}^2$. We generated a catalog of $5741$ sources above $5\sigma_{\text{rms}}$ from this image with a dynamic range of $21251$. We compared our catalog with the ASKAP, GLEAM-X, and TGSS surveys and provided complementary results on the source population and their spectral properties in the GAMA-23 field at low frequencies. The estimated normalized differential source counts lie within the range $0.5-4600$\,mJy. This demonstrates the sensitivity of uGMRT, even with limited observation time at 325\,MHz. We have compared our source counts with the ASKAP surveys at 887.5\,MHz and GLEAM-X at 200\,MHz as well as with previous uGMRT observations and found that this work is consistent with the ASKAP surveys, other observations, and simulations. \\
We have presented the characterization of dominant foreground, i.e., DGSE, in the GAMA-23 field with six different latitude target pointings. The results from this work show that the amplitude of DGSE lies within $1-20\,{\text{mK}^2}$ range as the GAMA-23 field is away from the zodiacal glow of the Galactic plane. The power law index varies within the range $1.5-2.4$. Overall, the GAMA-23 field has minimal foreground contamination and is favorable to study the 21 cm signal. In the future, we will analyze the overlapping region near the center of the GAMA-23 field in Band-3 and Band-2 ($120-250$\,MHz) to determine variations in the spectral and spatial behavior of DGSE in much more detail.

%% Also note that the akcnowlodgment environment does not support long amounts of text. If you have a lot of people and institutions to acknowledge, do not use this command. Instead, create a new 

\begin{acknowledgments}

R.S. and A.D. acknowledge the use of computational facilities at DAASE, IIT Indore, which were established through funding from the Department of Science and Technology (DST), Government of India, under the DST-FIST grant (SR/FST/PSII/2021/162 (C)). The authors thank the staff of the Giant Metrewave Radio Telescope (GMRT), operated by the National Centre for Radio Astrophysics (NCRA) of the Tata Institute of Fundamental Research (TIFR), for providing the observational support. R.S. acknowledges financial support from the UGC through the UGC-SRF fellowship. A.D. acknowledges support from the Science and Engineering Research Board (SERB), Department of Science and Technology, Government of India, under the Core Research Grant (CRG/2021/004025) for the project titled ``Observing the Cosmic Dawn in Multicolor using Next Generation Telescopes.'' 
\end{acknowledgments}


%\vspace{5mm}
%\facilities{uGMRT (NCRA-TIFR), UGC-JRF fellowship}


\software{astropy \citep{2018AJ....156..123A, 2022ApJ...935..167A}  
          %WsClean \citep{2014MNRAS.444..606O, 2017MNRAS.471..301O}, 
          %PyBDSF \citep{2015ascl.soft02007M}
          }



% \appendix

% \section{Appendix information}


\bibliography{sample631}{}
\bibliographystyle{aasjournal}



\end{document}